\documentclass[12pt]{iopart}

\usepackage{iopams}
\expandafter\let\csname equation*\endcsname\relax
\expandafter\let\csname endequation*\endcsname\relax
\usepackage{amsmath}
\usepackage{hyperref}
\usepackage{manuscriptStyleFileTia}
\usepackage[version=4]{mhchem}
\usepackage{orcidlink}
\usepackage{biblatex}
\newcommand{\bm}[1]{\mathbf{#1}}
\usepackage{graphicx}
\usepackage{siunitx}
\usepackage{lineno}

\date{\today}

\begin{document}

\title[Remote Sensing]{Latent Twins: A Framework for Scene Recognition and Fast Radiative Transfer Inversion in FORUM All-Sky Observations}

\author{Cristina Sgattoni$^{1,5}$\orcidlink{0000-0001-5734-0856}, Luca Sgheri$^2$\orcidlink{0000-0002-6014-9363},  Matthias Chung$^3$\orcidlink{0000-0001-7822-4539},  Michele Martinazzo$^4$\orcidlink{0000-0002-1440-0043}}

\address{$^1$Institute of BioEconomy (IBE), National Research Council (CNR), Via Madonna del Piano, 10, Sesto Fiorentino, 50019, Firenze, Italy.}

\address{$^2$Institute of Applied Mathematics
(IAC), National Research Council (CNR), Via Madonna del Piano, 10, Sesto Fiorentino, 50019, Firenze, Italy.}

\address{$^3$Department of Mathematics, Emory University, 400 Dowman Drive, Atlanta, GA, USA.}

\address{$^4$Department of Physics and Astronomy ’’Augusto Righi", University of Bologna, Viale Berti Pichat 6/2, Bologna, Italy.}

\address{$^5$INdAM Research Group GNCS, P.le Aldo Moro, 5, 00185 Roma, Italy.}

\eads{\mailto{cristina.sgattoni@cnr.it},  \mailto{luca.sgheri@cnr.it}, \mailto{matthias.chung@emory.edu}, \mailto{michele.martinazzo2@unibo.it}}

\vspace{10pt}
\begin{indented}
\item[]\today
\end{indented}

\begin{abstract}
The FORUM (Far-infrared Outgoing Radiation Understanding and Monitoring) mission will provide, for the first time, systematic far-infrared spectral measurements of Earth's outgoing radiation, enabling improved understanding of atmospheric processes and the radiation budget. Retrieving atmospheric states from these observations constitutes a high-dimensional, ill-posed inverse problem, particularly under cloudy-sky conditions where multiple-scattering effects are present. In this work, we develop a data-driven, physics-aware inversion framework for FORUM all-sky retrievals based on \emph{latent twins}: coupled autoencoders for atmospheric states and spectra, combined with bidirectional latent-space mappings. A lightweight model-consistency correction ensures physically plausible cloud variable reconstructions. The resulting framework demonstrates potential for retrievals of atmospheric, cloud and surface variables, providing information that can serve as a prior, initial guess, or surrogate for computationally expensive full-physics inversion methods. It also enables robust scene classification and near-instantaneous inference, making it suitable for operational near-real-time applications. We demonstrate its performance on synthetic FORUM-like data and discuss implications for future data assimilation and climate studies.
\end{abstract}

\vspace{2pc}
\noindent{\it Keywords}: deep learning, inverse problems, autoencoders, latent twins, remote sensing, atmospheric retrieval, scene recognition

\submitto{}

\section{Introduction and Background}\label{sec:intro}

FORUM (Far-infrared Outgoing Radiation Understanding and Monitoring) \cite{palchetti2020} is the ninth Earth Explorer mission selected by the European Space Agency in 2019 and scheduled to launch in 2027. The mission will employ a Fourier Transform Spectrometer to measure Earth's outgoing spectral radiances over the spectral range from 100 to 1600~\si{cm^{-1}}. This will provide, for the first time, systematic and spectrally resolved observations in the far-infrared (FIR) region (100-667~\si{cm^{-1}}) \cite{harries08}. FORUM's primary objective is to improve our understanding of Earth's infrared radiation budget and to characterize the role of FIR radiation in shaping the climate system.

Different components of the Earth system, including the surface, atmospheric gases, aerosols, and clouds, interact with electromagnetic radiation, producing distinct spectral signatures in the outgoing radiance. In the FIR, water vapor exhibits particularly strong absorption features, making FORUM measurements highly valuable for constraining atmospheric humidity and related climate processes \cite{Matricardi99}. In regions of low humidity, reduced water vapor absorption enables the retrieval of FIR surface emissivity from satellite observations \cite{ben2022}. FIR radiances are also strongly influenced by clouds and show greater sensitivity to ice particle scattering than those in the mid-infrared\cite{maestri03,saito2020,yang13}. FORUM measurements will therefore help to characterize the impact of clouds on the Earth's radiation budget, with a particular focus on high-level ice clouds.

The propagation and interaction of electromagnetic radiation through the atmosphere are described by the radiative transfer equation. Analytic solutions exist only for idealized, homogeneous, non-scattering media. In the real atmosphere, which exhibits strong vertical variability in temperature, composition, and cloud properties, radiative transfer must be solved numerically \cite{goody1989}. The task of predicting the radiation field from a given atmospheric state is referred to as the \emph{direct} or \emph{forward} problem.

In general, let $\mathbf{x} \in \mathbb{R}^n$ denote the atmospheric state,
and let $\mathbf{y} \in \mathbb{R}^q$ denote the corresponding radiance  then,
\begin{equation}\label{eq:forward}
    \mathbf{y} = \mathbf{F}^\to(\mathbf{x}) + \boldsymbol{\varepsilon},
\end{equation}
where $\mathbf{F}^\to : \mathbb{R}^n \to \mathbb{R}^q$ represents the composition of the radiative transfer calculations and instrumental effects, and $\boldsymbol{\varepsilon} \in \mathbb{R}^q$ denotes additive measurement noise. 
In this work, we are primarily interested in the \emph{inverse}, or \emph{retrieval}, problem, namely, reconstructing the atmospheric state $\mathbf{x}$ from measurements $\mathbf{y}$.

Inversion aims to infer unknown quantities of interest, $\mathbf{x} \in \mathbb{R}^n$, from observable data $\mathbf{y} \in \mathbb{R}^q$. Numerical inversion techniques, such as those based on Optimal Estimation (OE), are well established and widely used \cite{Rodgers1976,serio08,Turner2014,SERIO2024}. In a variational setting, the optimal state is obtained by minimizing a cost functional that balances data fidelity and regularization,
\begin{equation}\label{eq:variational}
    \min_{\mathbf{x} \in \mathcal{X}} \quad
        \mathcal{D}\bigl(\mathbf{F}^\to(\mathbf{x}), \mathbf{y}\bigr) + \mathcal{R}(\mathbf{x}),
\end{equation}
where $\mathcal{D}$ measures the discrepancy between the observations $\mathbf{y}$ and the model prediction $\mathbf{F}^\to(\mathbf{x})$, and $\mathcal{R}$ is a regularization term that penalizes implausible solutions \cite{tarantola2005inverse,engl1996regularization}. The forward operator can be expressed as the composition $\mathbf{F}^\to = \mathbf{p} \circ \mathbf{m}$, where $\mathbf{m}$ denotes the solution operator of the underlying physical model and $\mathbf{p}$ a projection onto the measurement space $\mathcal{Y}$.

While classical inversion techniques
remain powerful and widely used, they are often hindered by several limitations, mainly related to the need for careful hyperparameter tuning, sensitivity to problem setup, intrinsic ill-posedness, curse of dimensionality, and the substantial computational cost associated with repeated forward simulations \cite{kaipio2006statistical,calvetti2007introduction}. In remote sensing retrievals, these challenges become particularly severe due to the high dimensionality of the problem and time-consuming radiative transfer models.
For example, with the FORUM spectrometer expected to produce over \num{10000} spectra daily, and each full-physics forward simulation requiring several minutes, brute-force inversion is infeasible for near-real-time (NRT) applications as required for weather and climate prediction \cite{burton2023forum}. Current approaches aim to mitigate these limitations by accelerating the forward model. Notable examples include fast radiative transfer codes such as $\sigma$-IASI/F2N (Far-to-Near Infrared) \cite{amato2002,martinazzo2021,masiello2024b}, RTTOV \cite{matricardi04,saunders18} or PCRTM \cite{LiuXu}.

In response to these limitations, data-driven approaches have emerged as powerful alternatives. Rather than relying exclusively on explicit forward models, these methods take advantage of large-scale simulated and curated datasets to directly learn the mappings between measurements and unknown states. Deep neural networks, in particular, have demonstrated the ability to approximate highly nonlinear inverse operators and to accelerate retrieval by orders of magnitude compared with iterative variational schemes \cite{lecun2015deep,arridge2019solving}. For instance, autoencoders and related latent-variable models can be used to capture low-dimensional structure within high-dimensional data, providing surrogates that constrain reconstructions to physically plausible manifolds \cite{baldi1989neural,kramer1991nonlinear}. 

In previous work, we developed a data-driven inversion approach restricted to clear-sky conditions, combining a pseudoinverse with weighted regularization implemented via neural networks \cite{sgattoni2025}. This demonstrated the potential of neural surrogates for fast and physically informed retrievals. While effective as a proof of concept, the approach did not fully exploit the expressive power of modern nonlinear deep learning models, nor did it address the additional challenges posed by cloudy-sky conditions, where multiple scattering and radiative degeneracies make inversion particularly difficult. 

In this work, we tackle these challenges by developing a data-driven atmospheric retrieval framework based on \emph{latent twins} \cite{chung2025latent}, designed for fully all-sky applicability and practical use across the FORUM mission. Building on recent advances in latent twins and paired autoencoders for inverse problems \cite{chung2024likelihood,hart2025bayes,chung2025good}, we construct a data-driven yet physics-aware surrogate inversion framework for FORUM all-sky observations. The key idea is to learn coupled autoencoders for atmospheric states and spectra, together with (bi)-directional mappings between their latent spaces, so that the latent coupling acts as a surrogate forward and inverse operator, effectively creating a latent twin of the original forward and/or inverse map. On top of this latent surrogate, we introduce a lightweight model-consistency correction targeted at cloud-related variables, enforcing physically meaningful relationships among cloud microphysical properties.
We demonstrate that the resulting framework shows potential for the retrieval of atmospheric, cloud, and surface variables, providing information that can be used as a prior, initial guess, or surrogate for computationally expensive full-physics inversion methods. It also enables robust scene classification and, once trained, near-instantaneous inference, making it a promising candidate for operational near-real-time applications and future data assimilation efforts.

This paper is organized as follows. \Cref{sec:approach} presents the latent twins framework, detailing the coupled autoencoders, latent-space mappings, and the model-consistency corrections for cloud variables. 
In \Cref{sec:tsc}, we describe the construction of the training dataset, while \Cref{sec:do} details the organization of the atmospheric state and observation vectors.
Numerical results, computational costs, and technical details of the models, including network architectures, hyperparameters, and loss functions, are presented in \Cref{sec:numres}. 
This section is structured into three subsections. \Cref{sec:numres_retr} presents retrieval performance on atmospheric variables. \Cref{sec:numres_scene} covers scene classification, including clear, cloudy, and cloud-type identification, as well as the localization of cloud layers. Finally, \Cref{sec:numres_chi2} assesses the impact of the retrievals on reconstructed radiances via the forward model, including an additional correction for clear-sky cases.
We conclude with a discussion of implications and future directions in \Cref{sec:concl}.
Two appendices are also included. 
In \ref{sec:app1}, we review the radiative transfer model, including the treatment of multiple scattering, and describe the $\sigma$-IASI/F2N forward model used to generate training data.
\ref{sec:app2} describes the preprocessing of selected variables and how their transformation influences the reconstruction of radiances from the retrieved quantities.

\section{Latent Twin Radiative Transfer Inversion} \label{sec:approach}

Data-driven inversion techniques replace or augment classical iterative solvers by learning components of the inverse process directly from data. One class of methods trains a neural network to approximate the reconstruction operator $\mathbf{y} \mapsto \mathbf{x}$ from paired training data $\{(\mathbf{y},\mathbf{x})\}$, enabling fast inference but often requiring additional mechanisms to ensure robustness and data consistency \cite{arridge2019solving}. A second class, commonly referred to as \emph{unrolled} (or \emph{algorithm-unfolded}) methods, derives neural architectures by unrolling and truncating classical iterative schemes and learning iteration-dependent parameters or proximal operators, thereby combining interpretability with state-of-the-art performance \cite{gregor2010lista,adler2017iterative}. Alternative approaches focus on \emph{prior learning}. Plug-and-play methods incorporate powerful learned denoisers as implicit regularizers within model-based optimization frameworks \cite{afkham2021learning,venkatakrishnan2013pnp}, while generative models provide explicit low-dimensional priors by restricting reconstructions to the range of a trained generator \cite{bora2017generativecs}. More recently, diffusion and score-based generative models have emerged as flexible learned priors that can be combined with known measurement operators at test time, enabling sampling-based reconstructions and principled uncertainty quantification \cite{song2021scorebased}.

\paragraph{Latent twins framework.}
A promising recent direction in data-driven inversion is the development of \emph{latent twins} as surrogate models for forward and inverse operators. Realized as paired autoencoders, latent twins learn forward and inverse mappings through coupled latent representations, thereby improving stability and interpretability in ill-posed problems; see, for example, likelihood-free estimation \cite{chung2024likelihood}, Bayes-risk minimization \cite{hart2025bayes}, and the unifying perspective in \emph{Good Things Come in Pairs} \cite{chung2025good}.

\begin{figure}[H]
\begin{center}
\includegraphics[width=0.4
\textwidth]{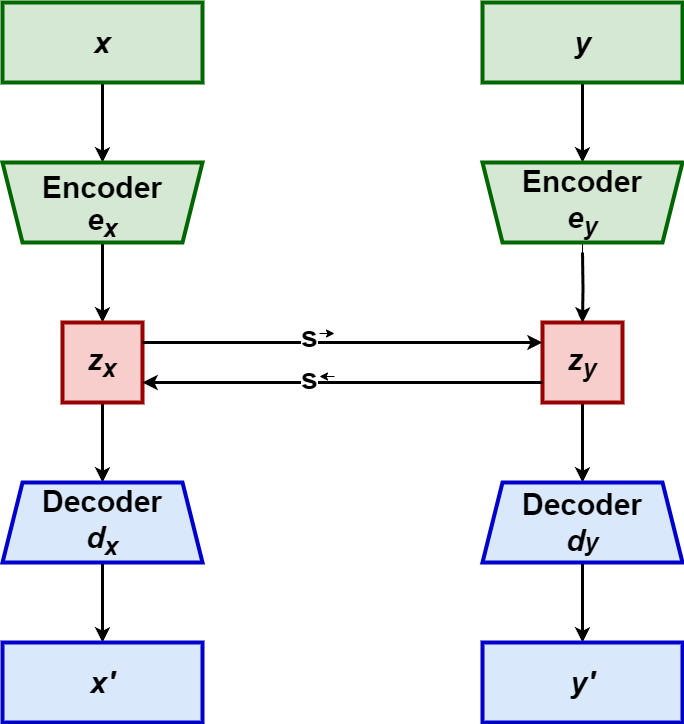}
\caption{Schematic of the Latent Twin (paired autoencoder) architecture for inverse problems. Two autoencoders are trained on the parameter space $\mathcal{X}$ and the observation space $\mathcal{Y}$, producing low-dimensional latent representations $z_x$ and $z_y$, respectively. The inverse operator is learned directly in latent space through a trainable mapping $s^{\gets} : \mathbf{z}_y \mapsto \mathbf{z}_x$, which is composed with the observation encoder and state decoder to form the full inverse surrogate/ latent twin $f^{\gets} = d_x \circ s^{\gets} \circ e_y$. By learning the inverse transformation in a shared latent coordinate system, the framework enables efficient and stable recovery of high-dimensional atmospheric states from observations, while preserving consistency between the data and model representations.
}\label{fig:pair}
\end{center}
\end{figure}

In this work, we adopt the latent twins framework as a data-driven yet physics-aware approach to atmospheric retrieval. Here, physics-awareness stems from restricting solutions to a learned physically plausible manifold and from additional model-consistency corrections introduced below. The central idea is to represent both the atmospheric state and the observed spectrum in low-dimensional latent spaces and to learn the forward and inverse relationships directly between these representations; see \Cref{fig:pair}. Let $\mathcal{X} \subseteq \mathbb{R}^n$ denote the space of atmospheric states and $\mathcal{Y} \subseteq \mathbb{R}^q$ the space of radiance spectra. We define two autoencoders: an atmospheric-state autoencoder $a_x = d_x \circ e_x$, mapping $\mathcal{X}$ to itself through a latent space $\mathcal{Z}_x$, and a spectral autoencoder $a_y = d_y \circ e_y$, mapping $\mathcal{Y}$ through a latent space $\mathcal{Z}_y$. The key component of latent twins is the introduction of \emph{trainable} mappings between $\mathcal{Z}_x$ and $\mathcal{Z}_y$. A latent forward map $s^\to : \mathcal{Z}_x \to \mathcal{Z}_y$ approximates the action of the radiative transfer operator in reduced coordinates, while a latent inverse map $s^\gets : \mathcal{Z}_y \to \mathcal{Z}_x$ serves as a latent retrieval operator. Combined with the decoders, these mappings define surrogate forward and inverse operators,
\[
f^\to(\mathbf{x}) = (d_y \circ s^\to \circ e_x)(\mathbf{x}),
\qquad
f^\gets(\mathbf{y}) = (d_x \circ s^\gets \circ e_y)(\mathbf{y}),
\]
which aim to approximate the true forward and inverse mappings on the data manifold, i.e., $f^\to \approx \mathbf{F}^\to$ and $f^\gets \approx \mathbf{F}^\gets$. In this work, our primary interest lies in retrieval, namely the construction of an accurate surrogate inverse operator $f^\gets$.

This separation between representation learning (autoencoders) and operator learning (latent mappings) is a distinguishing feature of latent twins, providing a transparent interpretation of how observations and states are related while yielding improved stability over direct surrogate inversion networks in ill-posed settings \cite{chung2024likelihood,hart2025bayes,chung2025good}.

An additional advantage of the latent twin construction is that it naturally induces a low-dimensional constraint on the inverse problem. In particular, $f^\gets(\mathbf{y})$ may be interpreted not only as a direct retrieval operator, but also as defining an implicit prior or admissible manifold for atmospheric states. By restricting reconstructions to the form $d_x(\mathbf{z}_x)$ with $\mathbf{z}_x \in \mathcal{Z}_x$, the inversion is confined to states representable by the learned atmospheric latent space, thereby suppressing nonphysical solutions that lie off the data manifold. From an inverse-problem perspective, the latent twin model may therefore be used as a standalone surrogate inversion, as a rapid initializer for physics-based retrieval, or as a regularizing reference; for example, in \Cref{eq:variational} one may incorporate the penalty
\[
\mathcal{R}(\mathbf{x}) = \lambda \, \|\mathbf{x} - f^\gets(\mathbf{y})\|_2^2, \qquad \lambda \geq 0.
\]

To support stable and accurate inversion, the training objective must jointly promote faithful reconstruction within each domain and consistent, well-aligned mappings between the corresponding latent spaces. The latent mapping may be trained in the forward direction, the inverse direction, or jointly in both directions, as adopted here.

\paragraph{Latent Twin training.}
Given representative parameter--observation pairs $\{(\mathbf{x}_j,\mathbf{y}_j)\}_{j=1}^J$, the autoencoders are trained using reconstruction losses
\begin{equation}
    \tfrac{1}{J} \sum_{j=1}^J \mathcal{J}_x\bigl((d_x \circ e_x)(\mathbf{x}_j), \mathbf{x}_j\bigr),
    \qquad \text{and} \qquad
    \tfrac{1}{J} \sum_{j=1}^J \mathcal{J}_y\bigl((d_y \circ e_y)(\mathbf{y}_j), \mathbf{y}_j\bigr),
\end{equation}
where $\mathcal{J}_x$ and $\mathcal{J}_y$ denote suitable discrepancy measures on the parameter and observation spaces, respectively.

For the latent mappings, a variety of loss formulations may be employed, including losses defined directly in latent space or full surrogate losses composed with the encoders and decoders \cite{chung2024likelihood}. In this work, we adopt the latter and train the latent mappings using full surrogate objectives,
\begin{equation}
    \tfrac{1}{J} \sum_{j=1}^J \mathcal{J}_{xy}\bigl(f^\to (\mathbf{x}_j), \mathbf{y}_j\bigr)
    \qquad \text{and} \qquad
    \tfrac{1}{J} \sum_{j=1}^J \mathcal{J}_{yx}\bigl(f^\gets(\mathbf{y}_j), \mathbf{x}_j\bigr),
\end{equation}
which enforce consistency between atmospheric states and observations through the learned forward and inverse surrogate operators.

We optimize all components jointly by minimizing a weighted combination of reconstruction and surrogate losses,
\begin{align}
\min_{\theta} \quad
\frac{1}{J} \sum_{j=1}^J \Big(
& \alpha_x \, \mathcal{J}_x\bigl((d_x \circ e_x)(\mathbf{x}_j), \mathbf{x}_j\bigr)
+ \alpha_y \, \mathcal{J}_y\bigl((d_y \circ e_y)(\mathbf{y}_j), \mathbf{y}_j\bigr) \nonumber \\
&\quad
+ \alpha_{xy} \, \mathcal{J}_{xy}\bigl(f^\to(\mathbf{x}_j), \mathbf{y}_j\bigr)
+ \alpha_{yx}  \mathcal{J}_{yx} \bigl( f^\gets(\mathbf{y}_j), \mathbf{x}_j\bigr)
\Big),
\label{eq:latent_twin_joint_loss}
\end{align}
where $\theta$ collects all trainable parameters of the encoders, decoders, and latent mappings, and $\alpha_x, \alpha_y, \alpha_{xy}, \alpha_{yx} > 0$ balance the individual loss terms. If only a forward or inverse surrogate is required, the corresponding mapping term may be omitted.

Empirically, we observe that joint optimization leads to better-aligned latent representations and simplifies the latent mappings, often allowing accurate surrogate operators to be represented by relatively low-complexity networks. In contrast to two-stage approaches that first train the autoencoders independently and then learn latent-space maps, joint training encourages the latent spaces to organize in a manner that is inherently compatible with the desired forward and/or inverse operator \cite{chung2024likelihood, hart2025bayes, chung2025good}.

\paragraph{Model-consistent inversion with latent twins.}
Purely data-driven surrogate reconstructions can deviate from physical consistency, particularly for cloud-related variables, which are among the most challenging atmospheric quantities to retrieve and are prone to cross-variable inconsistencies. To address this while retaining a fully data-driven workflow, we augment the latent twin framework with targeted consistency correction models that operate directly in physical variable space. For each variable requiring correction, an auxiliary neural network is trained independently on the same dataset as the latent twin and applied selectively during inference only when inconsistencies are detected, thereby avoiding the cost of a full-physics-based inversion while preserving efficiency and interpretability.

Formally, we introduce corrective maps of the form
$
g: \mathcal{X} \;\to\; \mathcal{X},$
trained as

\begin{align}
\min_{\theta_g} \quad
\frac{1}{J} \sum_{j=1}^J
\bigg(
& \gamma_{\mathrm{inc}} \,
\mathcal{J}_{\mathrm{inc}}\!\Bigl(
g\bigl(f^\gets(\mathbf{y}_j)\bigr), \mathbf{x}_j
\Bigr)
+
\gamma_{\mathrm{con}} \,
\mathcal{J}_{\mathrm{con}}\!\Bigl(
g\bigl(f^\gets(\mathbf{y}_j)\bigr), \mathbf{x}_j
\Bigr)
\nonumber \\
&\quad
+ \gamma_{\mathrm{p}} \,
\mathcal{J}_{\mathrm{p}}\!\Bigl(
g\bigl(f^\gets(\mathbf{y}_j)\bigr)
\Bigr)
\bigg),
\label{eq:corrective_loss_abstract}
\end{align}

where, $\theta_g$ collects all trainable parameters, $\mathcal{J}_{\mathrm{inc}}$ and $\mathcal{J}_{\mathrm{con}}$
measure discrepancies between corrected predictions and reference states over
complementary subsets of the state components, corresponding to inconsistent and
consistent elements, respectively.
The functional $\mathcal{J}_{\mathrm{p}}$ denotes an additional structural
regularization term enforcing desired internal consistency properties.
The positive weights $\gamma_{\mathrm{inc}}$, $\gamma_{\mathrm{con}}$,
$\gamma_{\mathrm{p}} > 0$ balance the contributions of the different loss terms.

\paragraph{Benefits for cloudy-sky retrievals.} For cloudy-sky conditions, where multiple scattering and high-dimensional parameterizations render full-physics inversion prohibitively slow, the latent twins framework offers several advantages. First, it potentially enables fast surrogate inversions that are suitable for near-real-time retrievals, a key requirement for operational remote sensing. Second, by constraining reconstructions to remain consistent with forward radiative transfer, the approach yields physics-informed solutions that avoid non-physical artifacts. Finally, the framework provides flexibility in training, as it naturally incorporates large-scaledatasets from multiple source, such as those derived from the fifth generation atmospheric reanalysis (ERA5) provided by the European Centre for Medium-Range Weather Forecasts (ECMWF) \cite{hersbach20} and simulated spectra from the $\sigma$-F2N forward model \cite{masiello2024b}, thereby ensuring robust generalization across diverse cloudy conditions. 
Taken together, these properties position latent twins as a powerful alternative to purely data-driven inversion methods, balancing computational efficiency with physical consistency.

\section{Dataset generation}\label{sec:tsc}
A major challenge in applying data-driven approaches to this remote sensing problem is the lack of an extensive dataset suitable for training neural networks. Consequently, a critical first step in the development and evaluation of the proposed method is the construction of a representative dataset through a simulation study. We next describe the generation of this dataset.

A dataset for the inversion problem is composed by a set of $J$ pairs $\{\bm{x}_j,\bm{y}_j\}_{j=1}^J.$ In our application, the vector $\bm{x}_j$ represents the atmospheric state, including atmospheric gases, surface properties such as Earth surface temperature and spectral emissivity, and cloud microphysical variables such as cloud mass content and cloud particle effective radius. The vector $\bm{y}_j$ denotes the simulated spectrum generated from $\bm{x}_j$ by applying the radiative transfer model, the convolution with the Instrument Spectral Response Function (ISRF), and the addition of noise according to the instrument variance-covariance matrix. This vector may also include ancillary information such as geolocation, acquisition time, and the atmospheric pressure levels.
Details of the underlying physical formulation of the radiative transfer model are provided in \ref{sec:app1}, along with relevant references, including \cite{masiello2024b, chou1999, tang2018}.

Meteorological data for this study were sourced from the ERA5 reanalysis dataset provided by ECMWF \cite{hersbach20}. Additional concentrations of atmospheric gases, including \ce{CO2}, \ce{CO}, \ce{CH4}, \ce{NO2}, \ce{HNO3}, and \ce{SO2}, were retrieved from the Copernicus Atmosphere Monitoring Service (CAMS) global Greenhouse Gas reanalysis dataset \cite{egg2020}. The remaining gas data were drawn from the Initial Guess 2 (IG2) database \cite{remedios07}. Surface emissivity was obtained from the Huang dataset \cite{huang2016} using a tailored approach that matches emissivity spectra, expressed as a function of wavenumber, to specific times and locations \cite{sgheri2024}, exploiting the Combined ASTER MODIS Emissivity over Land (CAMEL) database \cite{borbas2018,feltz2018}. The CAMEL data is available over land. On the sea, to provide some variability anyway to the emissivity profile, we applied the Masuda emissivity modification \cite{masuda1988} due to wind, using the Cross-Calibrated Multi-Platform Wind Vector Analysis Product \cite{wentz2015}. While the modification is minimal for nadir observations, it still prevents the generation of a large number of sample points with a constant emissivity profile, which could be easily detected by data-driven approaches.

\begin{figure}[H]
\begin{center}
\includegraphics[width=.49\textwidth]{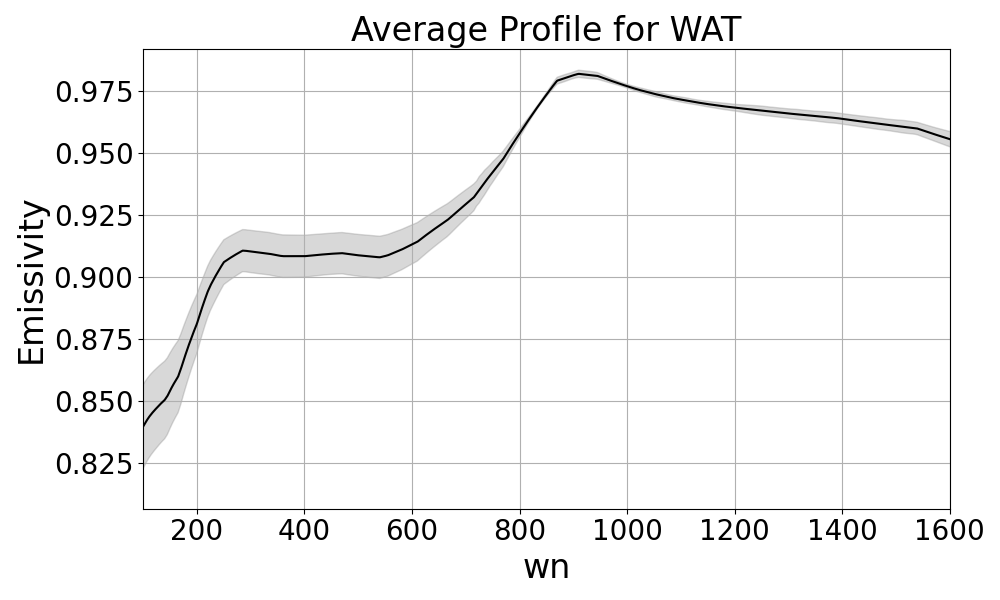}
\includegraphics[width=.49\textwidth]{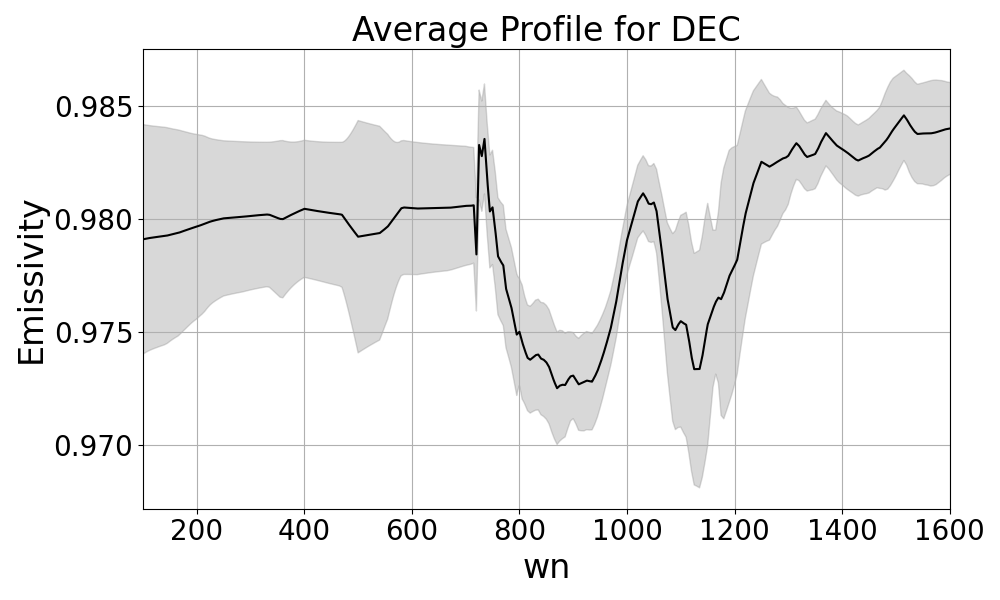}
\caption{Average January emissivity profiles with standard deviation (shaded) for ocean (WAT) points (left panel) and land points dominated by deciduous (DEC) land cover (right panel).}\label{fig:emiss}
\end{center}
\end{figure}

In the left panel of Figure~\ref{fig:emiss}, we show the average January emissivity profile over water, with the shaded region indicating the corresponding standard deviation. For comparison, the right panel displays the emissivity profiles for points whose fields of view are dominated by deciduous land cover. In this case, the variability reflects the application of the method proposed by \cite{sgheri2024}. 

For cloudy-sky scenes, the cloud ice water content and cloud specific liquid water content are taken from the ECMWF database. The average particle dimensions required for the multiple-scattering calculations are modeled using the Wyser model~\cite{wyser1998} for ice particles and the Martin model~\cite{martin1994} for liquid water droplets.

We define a scene to be cloudy if the total optical depth due to clouds at $900$~\si{cm^{-1}} exceeds $0.03$, a threshold consistent with cloud detectability, as supported by prior studies \cite{sgheri2022,sgattoni2024}.

The ECMWF database also provides a \textit{cloud fraction} parameter for each atmospheric pressure level, indicating the portion of the sky covered by clouds within the associated tile. To improve global cloud coverage representation, each cloud content could be rescaled according to its cloud fraction. This approach spreads the content over all the tile, so that thinner clouds for each dataset point are calculated. Since our focus is on cloudy conditions, we prefer to assume that the entire cloud content of each tile is concentrated at a single position. While this approach overestimates the number of cloudy scenes, it remains representative of actual cloud positioning and thickness.

Given the large volume of data required, a full-physics forward model such as the CLouds and Atmospheric Inversion Model (CLAIM), an advanced version of the inversion module of the FORUM End-to-End Simulator \cite{sgheri2022,sgheri2022a}, is computationally prohibitive. In particular, cloudy-sky forward simulations using the accelerated DIScrete Ordinate Radiative Transfer (DISORT) solver \cite{sgheri2018} can require several hours per case, depending on the cloud’s geometrical thickness and the characteristic particle size. Instead, we employ the $\sigma$-IASI/F2N code \cite{amato2002,maestri2021,martinazzo2021,masiello2024}, an extension of the $\sigma$-IASI framework that supports spectrometers operating across the infrared spectrum. The $\sigma$-IASI/F2N code parameterizes optical depths using low-order polynomials in temperature and, for water vapor, in the water-vapor concentration itself to account for self-broadening effects. Under cloudy-sky conditions, multiple-scattering effects are incorporated via the Chou scaling method \cite{chou1999}. Finally, a tailored preprocessing code, hereafter referred to as ERA5 Code, assembles the surface and atmospheric data and interpolates them onto the grids required for the forward simulations.

\begin{figure}[H]
\begin{center}
\includegraphics[width=1.
\textwidth]{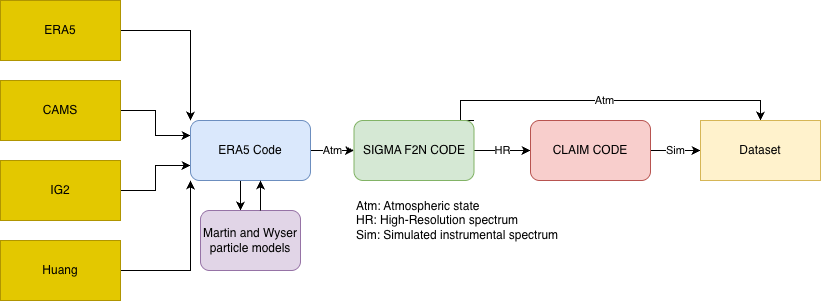}
\caption{Workflow of the algorithm for the generation of the dataset. Atmospheric input data are retrieved from source databases (yellow, left) and processed through models and preprocessing codes (blue and purple) before being ingested by the forward model (green). The resulting high-resolution spectra are then convolved and corrupted with noise (red) to generate the final simulated spectral database (yellow, right).}\label{fig:algorithm}
\end{center}
\end{figure}

The CLAIM code is used to perform the spectral convolution and to add instrument noise. For FORUM, we adopt a spectral resolution of $0.36$\si{cm^{-1}}, the Norton–Beer strong ISRF \cite{norton1976}, and Noise-Equivalent Spectral Radiance (NESR) levels taken from one of the PhaseA instrument studies \cite{sgheri2022}. The overall procedure used to generate the dataset is summarized in \Cref{fig:algorithm}.

\section{Data organization} \label{sec:do}

We define an atmospheric state (or scenario) as a collection of parameters describing atmospheric gas constituents, surface properties, and cloud characteristics that together fully determine the simulated radiance spectra. In this study, we vary a set of 11 key variables, while all other parameters are determined by the specific scene and are not treated as retrieval variables in the neural network architecture. These variables are: Earth’s surface temperature ($T^0$); 
air temperature ($\bm{T}$); water vapor ($\bm{w}_{\text{vap}}$); ozone ($\bm{o}$); surface spectral emissivity ($\bm{\varepsilon}$); cloud liquid water content ($\bm{c}_{\text{liq}}$); cloud ice water content ($\bm{c}_{\text{ice}}$); 
cloud liquid particle effective radius ($\bm{r}_{\text{liq}}$); cloud ice particle effective radius ($\bm{r}_{\text{ice}}$); and the liquid and ice cloud optical depths, $\tau_{\text{liq}}$ and $\tau_{\text{ice}}$, defined at $900~\si{cm^{-1}}$.

Among these parameters, Earth’s surface temperature is a scalar, whereas surface spectral emissivity is a wavenumber-dependent vector, typically represented using a linear spline. The remaining parameters are vertical profiles defined on atmospheric pressure layers. This vertical grid is fixed across the dataset; however, the lowest layers may be truncated to account for variations in Earth’s surface altitude with geolocation. As a result, the number of vertical layers associated with each parameter can vary from case to case, which poses a challenge when assembling all samples into a uniformly dimensioned input matrix for our method.

To address this issue and standardize the data, we adopt the approach outlined in \cite{sgattoni2025}, which can be interpreted as a form of stretching. Specifically, the pressure range of each atmospheric column is mapped onto a fixed set of $N=60$ standardized pressure layers, and all vertical profiles are interpolated onto this common grid. This procedure ensures that, although the physical pressure layers may vary between cases, the dimensionality of the resulting state representation remains constant. As a result, the atmospheric state vector $\bm{x}$ consists of $n=722$ components and is defined as
\[
\bm{x} =
\begin{bmatrix}
T^0 &
\bm{T} & \bm{w}_{\text{vap}} & \bm{o} &
\boldsymbol{\varepsilon} &
\bm{c}_{\text{liq}} & \bm{r}_{\text{liq}} & \bm{c}_{\text{ice}} &
 \bm{r}_{\text{ice}}
\end{bmatrix}^\top .
\]
Here, $T^0 \in \bbR$ and $\bm{T}$, $\bm{w}_{\text{vap}}$, $\bm{o}$, $\bm{c}_{\text{liq}}$, 
$\bm{r}_{\text{liq}}$,
$\bm{c}_{\text{ice}}$,  and $\bm{r}_{\text{ice}} \in \bbR^{60}$. The cloud-related variables $\bm{c}_{\text{liq}}$,
$\bm{r}_{\text{liq}}$,$\bm{c}_{\text{ice}}$,  and $\bm{r}_{\text{ice}}$ are identically zero under clear-sky conditions or in atmospheric layers without clouds. The surface spectral emissivity $\boldsymbol{\varepsilon} \in \bbR^{301}$ is defined over the spectral interval from $100$ to $\num{1600}~\si{cm^{-1}}$ and is interpolated onto a regular grid with $5~\si{cm^{-1}}$ spacing. Finally, the water vapor and ozone profiles are represented in a transformed space, as detailed in \ref{sec:app2}, where the motivation and precise formulation are provided.

The cloud optical depths, $\tau_{\text{liq}}$ and $\tau_{\text{ice}}$, are correlated quantities that can be computed, respectively, from the pairs $(\bm{c}_{\text{liq}}, \bm{r}_{\text{liq}})$ and $(\bm{c}_{\text{ice}}, \bm{r}_{\text{ice}})$ using standard formulations \cite{yang2001,yang2003}. For ice clouds, the optical depth is given by
\begin{equation} \label{eq:tau}
    \tau_{\text{ice}}
    = \frac{3}{2} \sum_i
    \frac{Q_{\text{ext}}^i \, c_{\text{ice}}^i \, \rho_{\text{air}}}
         {r_{\text{ice}}^i \, \rho_{\text{ice}}}
    \, \Delta z^i ,
\end{equation}
where $c_{\text{ice}}^i$ and $r_{\text{ice}}^i$ denote the cloud ice water content and cloud ice particle effective radius in the $i$-th atmospheric layer, respectively; $\rho_{\text{air}}$ and $\rho_{\text{ice}}$ are the air and ice densities; $\Delta z^i$ is the geometric thickness of layer $i$; and $Q_{\text{ext}}^i$ is the extinction efficiency \cite{stamnes1988,turner2003}. The latter depends on the cloud particle effective radius $r_{\text{ice}}^i$, the target wavenumber (here $900~\si{cm^{-1}}$), and the assumed cloud microphysical properties. An analogous expression is used to compute the liquid cloud optical depth $\tau_{\text{liq}}$.

The quantities $\tau_{\text{liq}}$ and $\tau_{\text{ice}}$ provide integrated measures of the optical thickness of liquid and ice clouds, respectively. Their sum, $\tau = \tau_{\text{liq}} + \tau_{\text{ice}}$, is used, for example, to support scene classification into clear- or cloudy-sky conditions. Although these variables are not directly included as inputs to our learning architecture, they are computed analytically at a later stage for diagnostic and classification purposes.

The radiance-related variables $\bm{y}$ consist of $q=\num{4233}$ components and are defined as
\begin{equation}
\bm{y} =
\begin{bmatrix}
y_{\text{lon}} &
y_{\text{lat}} &
d_{\text{loc}} &
h_{\text{loc}} &
\bm{p} &
\bm{s}
\end{bmatrix}^\top .
\end{equation}
These include the case location, specified by longitude ($y_{\text{lon}} \in \bbR$) and latitude ($y_{\text{lat}} \in \bbR$); the time of acquisition, given by the month ($d_{\text{loc}} \in \bbN$) and the local solar time ($h_{\text{loc}} \in \bbR$); the pressure-layer vector $\bm{p} \in \bbR^{60}$; and the simulated FORUM noisy radiance spectrum $\bm{s} \in \bbR^{\num{4169}}$. Longitude values range from $-178^\circ$ to $178^\circ$, and latitude values from $-88^\circ$ to $88^\circ$, both sampled in steps of $2^\circ$. All measurements are assumed to be acquired at the same UTC time. Consequently, the local solar time, expressed in decimal hours, is fully determined by the longitude and can be computed as $h_{\text{loc}} = 12 + \frac{12}{180}\, y_{\text{lon}}$.

\begin{figure}[H]
\begin{center}
\includegraphics[width=1.
\textwidth]{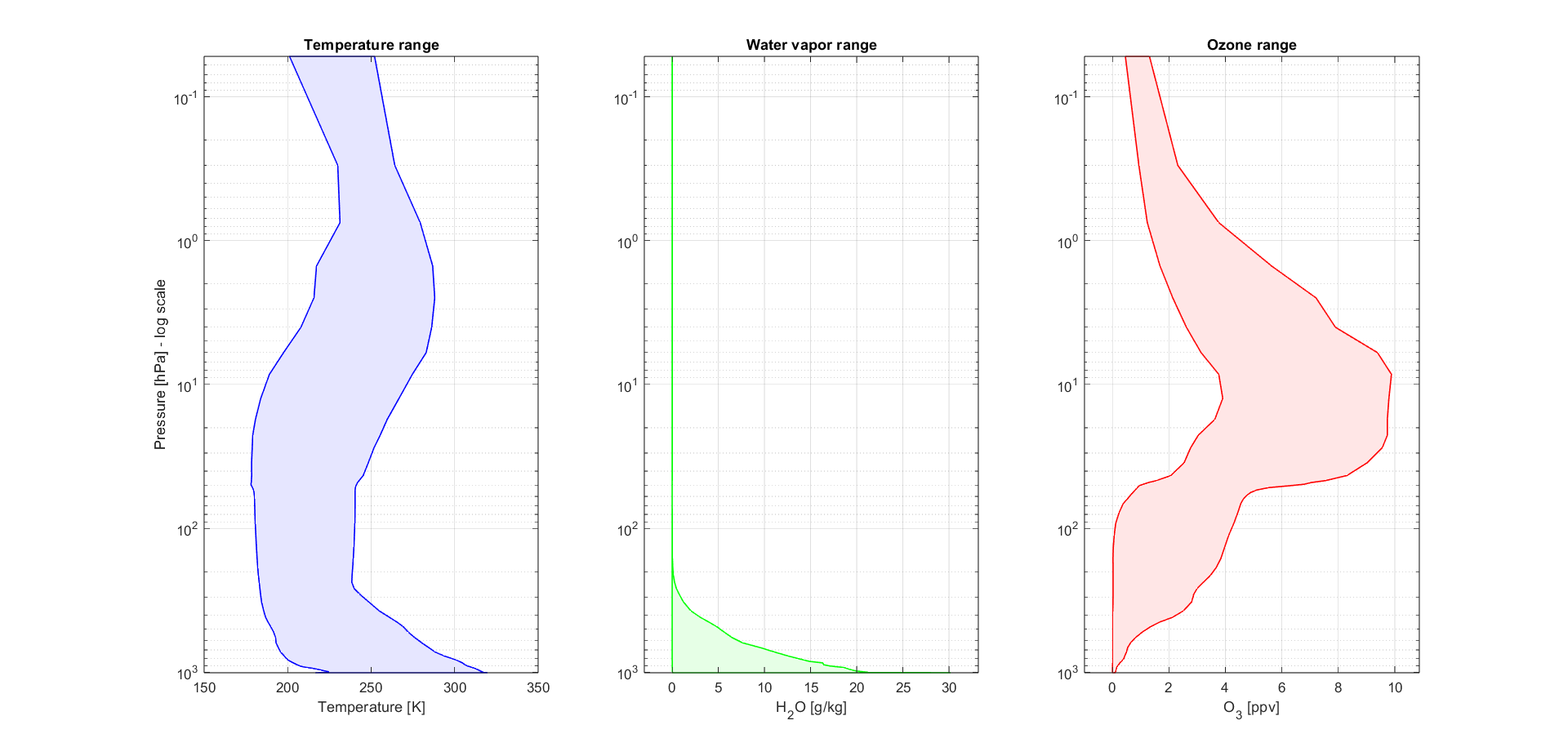}
\caption{Variables associated with atmospheric scenarios, presented with their ranges across pressure layers. The lines indicate the minimum and maximum values in the database, while the shaded area shows the variability between them. }\label{fig:inputvar_range}
\end{center}
\end{figure}

In our model configuration, the pressure grid comprises 60 atmospheric layers. Since the surface pressure $p^{0} \in \bbR$ is provided, we replace  the lowest pressure layer $p^{1}$ with $p^{0}$ and interpolate all remaining variables accordingly. The spectra are provided across $\num{4169}$ spectral points within the wavenumbers range from 100 \si{cm^{-1}} to \num{1600.5} \si{cm^{-1}} with steps of $0.36$ \si{cm^{-1}} and are generated using the code architecture described in \Cref{sec:tsc} giving as an input the corresponding atmospheric scenarios.
In Tables \ref{tab:inputvar}, \ref{tab:outputvar}, and \ref{tab:locvar}, a detailed and comprehensive description of all the variables involved is provided, categorized into atmospheric scenario variables, measurement variables, and location description variables, respectively. For each variable, we report the notation, description, unit of measurement, and the range of values spanned by the entire dataset. 
The values of $\bm{p}, \bm{T}, \bm{w}_{\text{vap}}$ and $\bm{o}$,  vary significantly across layers, so their ranges are shown in \Cref{fig:inputvar_range}.

\begin{table}[h]
    \centering
    \begin{tabular}{|c|c|c|c|}\hline
        $T^{0}$ & Earth surface temperature & \si{K} & [206.55, 338.32]
        \\
        $T^{i}$ & air temperature at layer $i$ & \si{K} & Figure \ref{fig:inputvar_range} 
        \\ 
        $w^{i}_{\text{vap}}$ & water vapor concentration at layer $i$ & \si{g/kg} & 
        Figure \ref{fig:inputvar_range} 
        \\ 
        $o^i$ & ozone concentration at layer $i$ & \si{ppv} & Figure \ref{fig:inputvar_range} 
        \\ 
        $\varepsilon^k$ & surface spectral emissivity at wavenumber $k$ & \si{1} & [0.84, 1]
        \\ 
        $c_{\text{liq}}^i$ & cloud liquid water content in layer $i$ & \si{kg/kg}  &  [0, $9\times10^{-4}$]
        \\
        $c_{\text{ice}}^i$ & cloud ice water content in layer $i$ & \si{kg/kg} & [0, $5\times10^{-4}$]
        \\
        $r_{\text{liq}}^i$ & cloud liquid particle effective radius in layer $i$ & \si{\mu m} & [0, 10.32]
        \\
        $r_{\text{ice}}^i$ & cloud ice particle effective radius in layer $i$ & \si{\mu m} & [0, 249.8]
        \\
        $\tau_{\text{liq}}$ & liquid cloud optical depth  & \si{1} & N/A 
        \\
        $\tau_{\text{ice}}$ & ice cloud optical depth & \si{1} & N/A 
        \\  \hline
    \end{tabular}
    \caption{Variables associated with the atmospheric scenarios, presented with their notation, description, unit of measurement and corresponding value ranges.}
    \label{tab:inputvar}
\end{table}

\begin{table}[h]
    \centering
    \begin{tabular}{|c|c|c|c|}\hline
        $s^k$ & radiance at wavenumber $k$ & \si{W /(m^2 \cdot\, sr \cdot cm^{-1})} & N/A 
        \\ \hline
    \end{tabular}
    \caption{Variables associated with the measurements, presented with their notation, description and unit of measurement and corresponding value ranges.}
    \label{tab:outputvar}
\end{table}

\begin{table}[h]
    \centering
    \begin{tabular}{|c|c|c|c|} \hline
        $p^{0}$ & Earth surface pressure & \si{hPa} & [519.98, 1038.52]
        \\
        $p^i$ & air pressure at layer $i$ & \si{hPa}  & Figure \ref{fig:inputvar_range} 
        \\ 
        $y_{\text{lon}}$ & longitudinal position & degrees (\si{\circ}) & $-178$ to $178$ in steps of 2  
        \\ 
        $y_{\text{lat}}$ & latitudinal position & degrees (\si{\circ}) & $-88$ to 88 in steps of 2  
        \\ 
        $d_{\text{loc}}$ & month & 1 & \{1,7\} 
        \\
        $h_{\text{loc}}$ & local time & 1 & 
       [0.1, 23.9].  \\ \hline
    \end{tabular}
    \caption{Variables related to the measurement locations, including their notation, description and unit of measurement and corresponding value ranges.}
    \label{tab:locvar}
\end{table}

For our analysis, we collect atmospheric scenario data from January and July 2021 at 12:00 UTC, covering the entire globe using a grid with $2^\circ$ increments in both latitude and longitude for a total amount of $J_{\text{all}}=\num{31862}$ cases. 



To train and evaluate the models, the dataset is partitioned into training and test sets. Of the total $J_{\text{all}}=\num{31862}$ cases, $J_{\text{train}}=\num{27000}$ are randomly assigned to the training set (approximately $85\%$), while the remaining $J_{\text{test}}=\num{4862}$ cases (approximately $15\%$) form the test set. Prior to training, all features are normalized using min–max scaling to map their values to the interval $[0,1]$. To avoid data leakage, the test set is normalized using the same minimum and maximum values computed from the training set, ensuring that the model is evaluated on previously unseen data. Some test samples may therefore fall slightly outside the interval $[0,1]$ after normalization; however, this minor overflow is naturally handled by the network’s activation functions.

\section{Numerical results} \label{sec:numres}

In this section, we present the numerical results obtained with our proposed approach. We first describe the training setup, including the loss functions, network architectures, and hyperparameter configurations used throughout our experiments in \Cref{sec:ltdesign}. Section \ref{sec:numres_retr} examines the performance of the retrieval of atmospheric variables, including an additional analysis focused on cloud-related variables, which are generally more challenging to retrieve. Section \ref{sec:numres_scene} evaluates the network’s ability to discriminate the scene type, distinguishing clear from cloudy conditions, assessing cloud type, and identifying cloud position. Finally, Section \ref{sec:numres_chi2} provides a statistical analysis of the radiance reconstruction, based on the $\chi^2$ metric, computed by applying the forward model to the retrieved atmospheric variables.

\subsection{Latent twin design}\label{sec:ltdesign}

\paragraph{Network architecture.} For simplicity, we employ standard multilayer perceptron (MLP) architectures for all neural network components of the latent twin. As explained in Section~\ref{sec:approach}, each autoencoder consists of an encoder and a decoder. The encoder is composed of fully connected layers with ReLU (Rectified Linear Unit) activation functions, progressively reducing the input dimensionality to a latent space of size $n_{\text{latent}}$. The decoder first maps the latent representation through a linear layer of the same dimensionality and then mirrors the encoder structure to reconstruct the input from the latent variables. An additional ReLU activation is applied after the final decoder layer to enforce non-negativity of the reconstructions.

For the autoencoder on $\mathbf{x}$, the encoder $e_x$ comprises two fully connected layers. 
The first hidden layer has size $(n + n_\text{latent}^x)/2$ = 617, and the second layer maps to the latent space of size $n_\text{latent}^x = 512$.  The decoder mirrors this structure with two layers of sizes $n_\text{latent}^x$ and $(n + n_\text{latent}^x)/2$, before reconstructing the original input. Similarly, the second autoencoder on $\mathbf{y}$ uses an encoder with two hidden layers of sizes $\lfloor (q + q_\text{latent}^y)/2 \rfloor= \num{2372}$ and $q_\text{latent}^y = 512$, mirrored by a decoder with the same layer sizes in reverse order. 


Trainable linear mappings are introduced between the latent spaces of the two autoencoders to model the forward ($s^{\to} :\bm{z}_x \mapsto \bm{z}_y$) and inverse ($s^{\gets} :\bm{z}_y \mapsto \bm{z}_x$) relationships. Accordingly, these mappings are implemented as single fully connected layers without activation functions. They enable the propagation of latent representations between the two domains, yielding surrogate latent representations $f^\to$ and $f^\gets$ for both the forward and inverse processes.

\paragraph{Training of the latent twin.}
The latent twin architecture is trained jointly by minimizing a composite loss function consisting of four terms, as defined in \Cref{eq:latent_twin_joint_loss}. Each term is formulated using the mean-squared error (MSE),
\[
\mathcal{J}(\mathbf{a}, \mathbf{b}) = \lVert \mathbf{a} - \mathbf{b} \rVert_2^2,
\]
and measures either reconstruction fidelity or cross-domain consistency in latent space. Specifically, the loss comprises: (i) reconstruction losses for the atmospheric-state and radiance autoencoders; (ii) a forward surrogate loss comparing decoded radiances obtained by mapping atmospheric latent variables into the radiance latent space; and (iii) an inverse surrogate loss comparing decoded atmospheric states obtained by mapping radiance latent variables into the atmospheric latent space. The individual loss terms are combined linearly with weights
\[
\alpha_x = \alpha_y = \alpha_{yx} = 1, 
\qquad
\alpha_{xy} = 0.05,
\]
reflecting our primary emphasis on accurate inversion from radiance observations to atmospheric states rather than on forward radiance prediction. This asymmetric weighting encourages faithful reconstruction of atmospheric variables while maintaining sufficient coupling between the two latent spaces to ensure stable and physically meaningful representations.

All models are implemented in PyTorch and trained using mini-batch optimization with a batch size of 512. Training is conducted over \num{10000} epochs, where each epoch corresponds to a full pass through the training dataset. Network parameters are optimized using the Adam optimizer with an initial learning rate of $\ell_r = 10^{-3}$. A step-based learning rate scheduler is applied, reducing the learning rate by a factor of $0.75$ every 500 training iterations. Training data are randomly shuffled at the beginning of each epoch to improve generalization and prevent the network from learning spurious correlations induced by sample ordering, while test data are processed without shuffling to ensure reproducibility. Training is carried out on a GPU-enabled Apple MacBook Pro with an M1 processor and 16\,GB of memory with training times of about 12 hours.


\paragraph{Additional training}
For certain variables, particularly those related to cloud properties, an additional training phase is required to enforce physical consistency, as discussed in Section~\ref{sec:approach}.

In some retrievals, for example, the inferred cloud particle effective radius is nonzero in atmospheric layers where the corresponding cloud content is zero, or vice versa, which is physically inconsistent. Although this behavior occurs only in a limited subset of cases, it motivates the introduction of an additional correction stage designed to enforce physical consistency among cloud microphysical variables. In the following, we describe the configuration of this post-processing stage, including the network architecture, loss formulation, and training procedure.

The corrective mechanism is implemented using two small, independent multilayer perceptrons: one for liquid-phase variables (cloud liquid water content and cloud liquid particle effective radius) and one for ice-phase variables (cloud ice water content and cloud ice particle effective radius). Each network consists of three fully connected hidden layers with 4, 3, and 2 neurons, respectively, and employs Softplus activation functions throughout. The input to each network comprises five predictors: the preliminary retrievals of cloud content and cloud particle effective radius, the retrieved temperature and water vapor, and the true atmospheric pressure. These predictors are chosen because cloud content and particle size are primarily controlled by local thermodynamic conditions and humidity. The networks output corrected, physically consistent estimates of the cloud microphysical variables.

The corrective networks are trained independently, while all parameters of the main autoencoders and latent mapping networks are kept frozen. Training is performed on the full training set. During inference, the corrective models are applied only to atmospheric layers in which physical inconsistencies are detected after the first-stage retrieval, leaving already consistent layers unchanged.

The corrective networks are trained by minimizing a composite loss function consisting of three terms, as defined in \Cref{eq:corrective_loss_abstract}.
The first two terms are based on the MSE between predicted and reference values,
\[
\mathcal{J}_{\mathrm{inc}}(\mathbf{a}, \mathbf{b}) = \lVert \mathbf{a} - \mathbf{b} \rVert_2^2, \quad 
\mathcal{J}_{\mathrm{con}}(\mathbf{a}, \mathbf{b}) = \lVert \mathbf{a} - \mathbf{b} \rVert_2^2,
\]
measuring the reconstruction accuracy separately over atmospheric layers identified as physically inconsistent and consistent after the first-stage retrieval.
These terms are weighted by coefficients $\gamma_{\text{inc}}$ and $\gamma_{\text{con}}$, respectively, allowing differential emphasis on correcting inconsistent retrievals while preserving already consistent ones.

The third term is a penalty term, weighted by $\gamma_{\text{p}}$, and is formulated using indicator functions discouraging discrepancies between the
zero/nonzero states of cloud content and cloud particle effective radius:

\[
\mathcal{J}_{\mathrm{p}}(\mathbf{a}, \mathbf{b}) = 
\big\| \mathbf{1}_{\{\mathbf{a} = 0\}} - \mathbf{1}_{\{\mathbf{b} = 0\}} \big\|_1.
\]

This term is applied only to layers identified as inconsistent and is normalized by the number of such layers for the liquid and ice phases, respectively.

Together, these loss acknowledging components encourage accurate reconstruction of cloud microphysical variables while explicitly enforcing physically consistent relationships between cloud presence and particle size.

Training is carried out using mini-batches of size $512$ and the Adam optimizer, with an initial learning rate of $10^{-4}$ and a weight decay of $10^{-5}$ for regularization. A learning rate scheduler reduces the learning rate by a factor of $0.5$ when the validation loss does not improve for five consecutive epochs. The networks are trained for $500$ epochs with $\gamma_{\text{con}} = 1$ and $\gamma_{\text{inc}} = \gamma_{\text{p}} = 5$, thereby emphasizing the correction of physically inconsistent predictions. The weighting scheme is subsequently reversed for an additional $500$ epochs to reinforce accuracy on physically consistent layers while preserving overall physical coherence.

At initialization, the training set contains $\num{409705}$ inconsistent layers, with $\num{71822}$ inconsistencies in the test set. After training, all such inconsistencies are fully resolved. The total loss decreases from $5.41$ at initialization to $0.04$ at convergence. Owing to the small size of the corrective networks, this post-processing stage is computationally inexpensive and can be efficiently executed on standard CPU hardware, without requiring GPU acceleration.

\subsection{Retrieval errors on atmospheric variables}
\label{sec:numres_retr}

This subsection assesses the performance of the trained latent twin on the test set, consisting of $J_{\text{test}}=\num{4862}$ cases. Given a test sample $\mathbf{y}_j$ with corresponding reference atmospheric state $\mathbf{x}_j$ and its prediction $\hat{\mathbf{x}}_j = f^{\gets}(\mathbf{y}_j)$, we define the residual vector as
\begin{equation}
\boldsymbol{\xi}_j = \mathbf{x}_j - f^{\gets}(\mathbf{y}_j) .
\end{equation}
Over the full test set, we evaluate two error metrics: the \textit{mean bias error} (\textbf{MBE}), which captures systematic over- or underestimation, and the \textit{mean absolute error} (\textbf{MAE}), which quantifies the average magnitude of deviations irrespective of sign,
\begin{equation}
\textbf{MBE} = \frac{1}{J_{\text{test}}} \sum_{j=1}^{J_{\text{test}}} \boldsymbol{\xi}_j,
\qquad
\textbf{MAE} = \frac{1}{J_{\text{test}}} \sum_{j=1}^{J_{\text{test}}} \lvert \boldsymbol{\xi}_j \rvert .
\end{equation}
These metrics are consistent with those previously employed for the clear-sky retrieval architecture in \cite{sgattoni2025}.

For the results presented in this section, and to enable consistent aggregation across the entire test set, all atmospheric profiles are interpolated onto a common linear pressure grid consisting of 100 layers, spanning pressures from $\num{1054}~\si{hPa}$ to $1~\si{hPa}$. This interpolation is used exclusively for computing and visualizing the \textbf{MBE} and \textbf{MAE} profiles. Individual-case plots, however, retain the original 60-layer pressure grids, which vary between cases.

\begin{figure}[H]
\centering
\includegraphics[width=\textwidth]{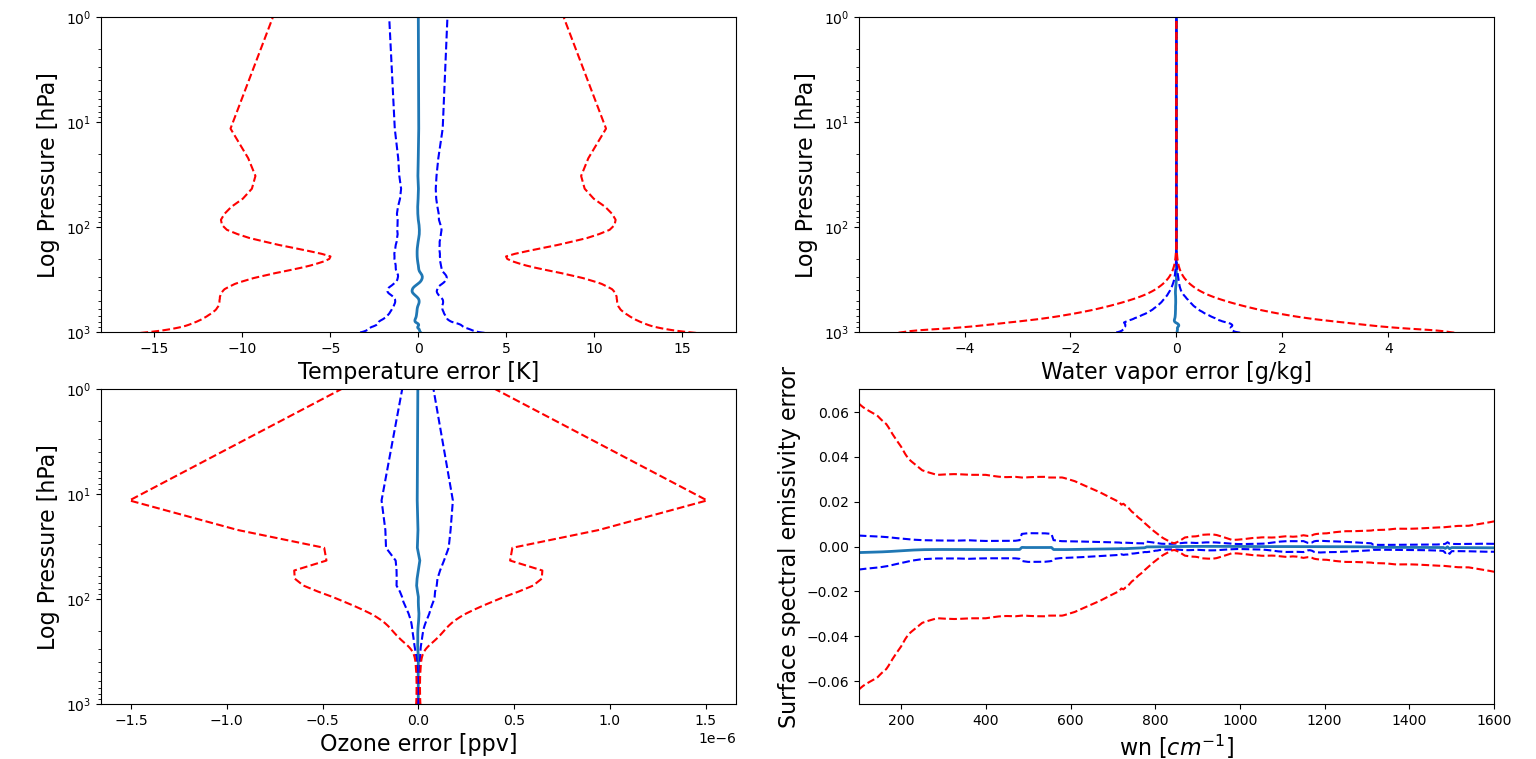}
\caption{Retrieval results for atmospheric state variables: temperature, water vapor, ozone, and spectral emissivity (clockwise from top-left). The blue solid line shows the \textbf{MBE}, while the blue and red dashed lines indicate $\textbf{MBE} \pm \textbf{MAE}$ and $\pm\,\textbf{MAD}$, respectively. For surface temperature, the mean bias error is $0.035~\si{K}$, the mean absolute error is $2.649~\si{K}$, and the mean absolute deviation of the scenes is $17.413~\si{K}$.}
\label{fig:avg1}
\end{figure}

Figure~\ref{fig:avg1} illustrates the retrieval results for selected atmospheric state variables: temperature, water vapor, ozone, and surface spectral emissivity (arranged clockwise). In each panel, the blue solid line denotes the \textbf{MBE}, while the blue dashed lines indicate the interval defined by $\textbf{MBE} \pm \textbf{MAE}$. The red dashed lines represent the intrinsic variability of the true atmospheric state, quantified as $\pm\,\textbf{MAD}$, where the mean absolute deviation (MAD) is defined as
\begin{equation}
\textbf{MAD} = \frac{1}{J_{\text{test}}} \sum_{j=1}^{J_{\text{test}}} \lvert \bm{x}_{j} - \bm{\bar{x}} \rvert ,
\end{equation}
and $\bm{\bar{x}}$ denotes the mean atmospheric state computed over all test samples. Results for the surface temperature are reported separately in the figure caption.

\begin{figure}[H]
\centering
\includegraphics[width=\textwidth]{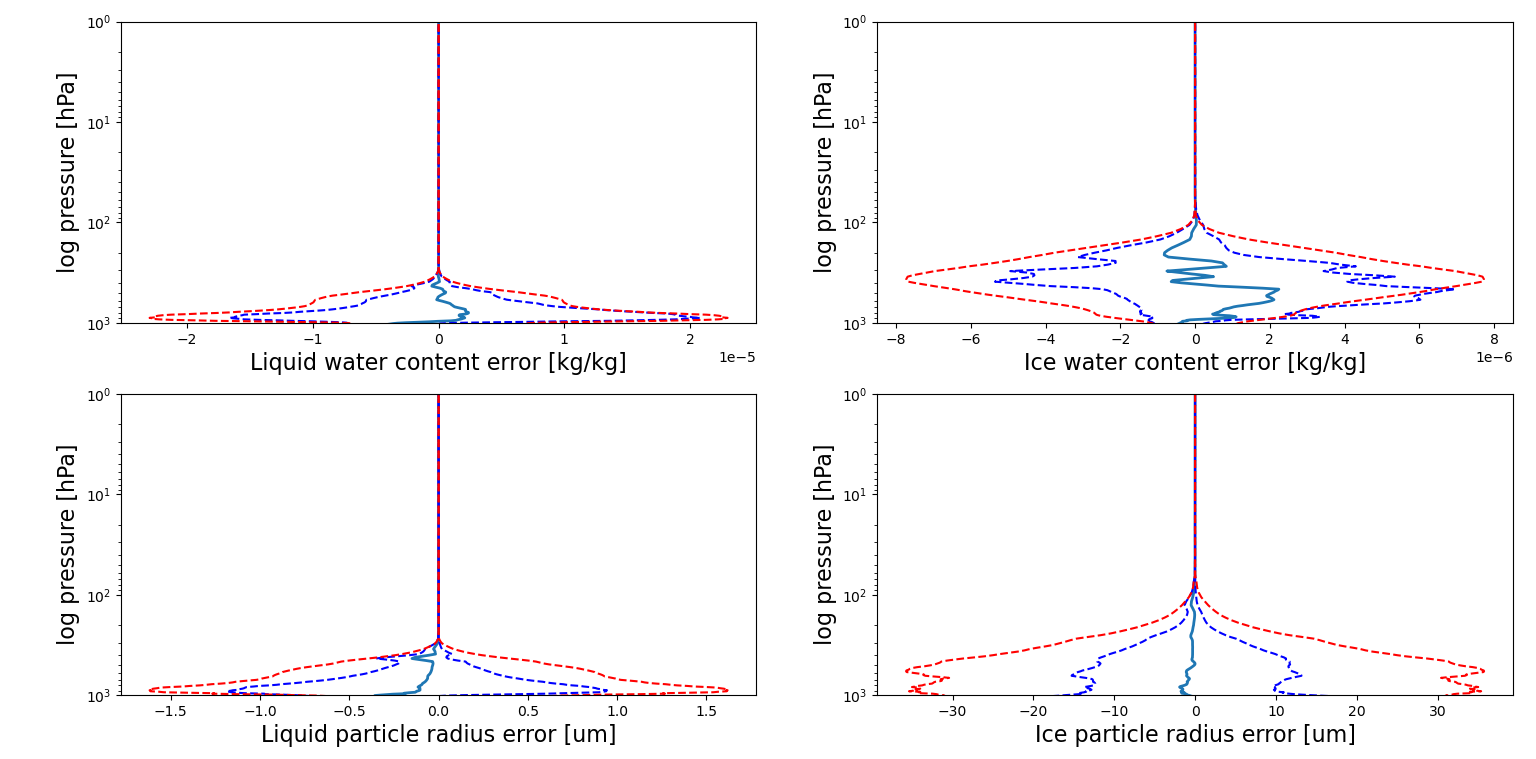}
\caption{Retrieval results for cloud-related parameters: cloud liquid water content, cloud ice water content, cloud liquid particle effective radius, and cloud ice particle effective radius (clockwise from top-left). The solid lines represent the \textbf{MBE}, while the dashed blue and red lines indicate $\textbf{MBE} \pm \textbf{MAE}$ and $\pm\,\textbf{MAD}$, respectively.}
\label{fig:avgcloud1}
\end{figure}

Figure~\ref{fig:avgcloud1} presents the same analysis for the cloud-related parameters: cloud liquid water content, cloud ice water content, cloud liquid particle effective radius, and cloud ice particle effective radius (arranged clockwise). As before, solid lines correspond to the \textbf{MBE}, while blue and red dashed lines indicate $\textbf{MBE} \pm \textbf{MAE}$ and $\pm\,\textbf{MAD}$, respectively. This representation simultaneously conveys the retrieval bias and an envelope of the expected absolute error across the test set, relative to the intrinsic variability of the atmospheric state.

\begin{figure}[H]
\centering
\includegraphics[width=\textwidth]{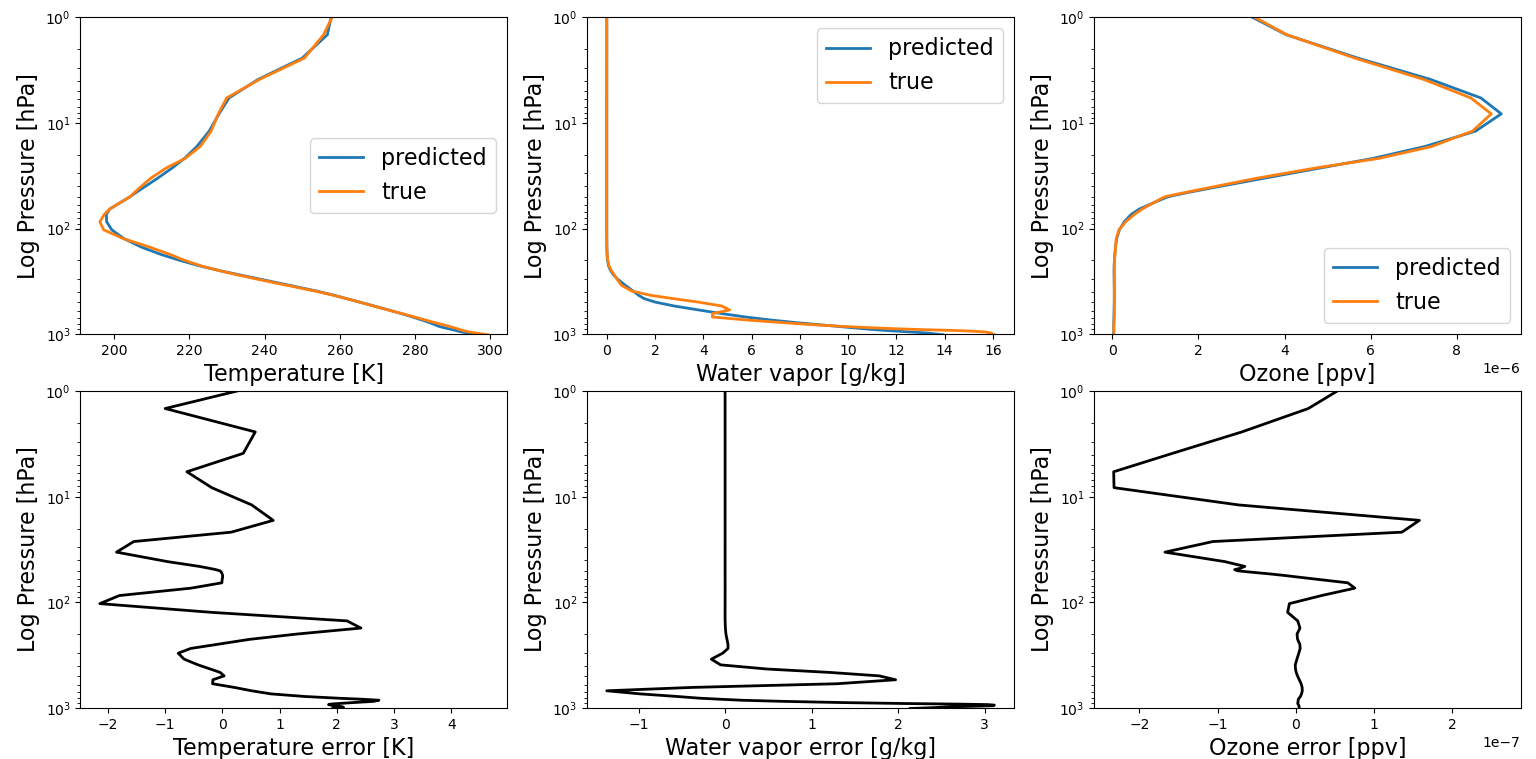}
\caption{Retrieved atmospheric profiles for a randomly selected test case (case number~3102). The first row shows the retrieved profiles (blue) compared to the reference profiles (orange), while the second row displays the corresponding residuals (black). For surface temperature, the predicted value is $299~\si{K}$ and the reference value is $301~\si{K}$.}
\label{fig:case5atm}
\end{figure}

\begin{figure}[H]
\centering
\includegraphics[width=0.8\textwidth]{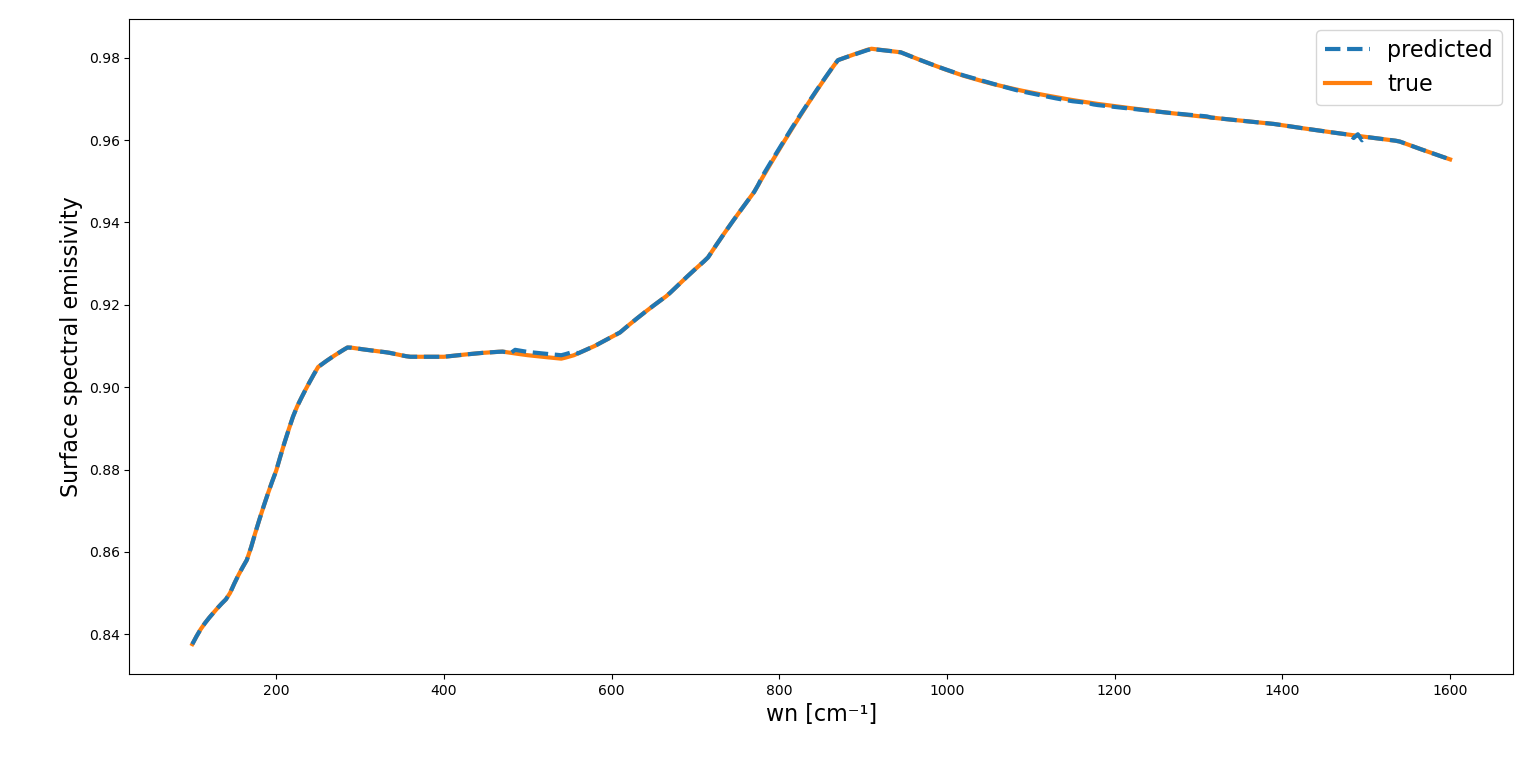}
\caption{Retrieved spectral emissivity for the same randomly selected test case (case number~3102). The retrieved values are shown as a dashed blue line, while the reference values are shown in orange.}
\label{fig:case5emi}
\end{figure}

\begin{figure}[H]
\begin{center}
\includegraphics[width=1.
\textwidth]{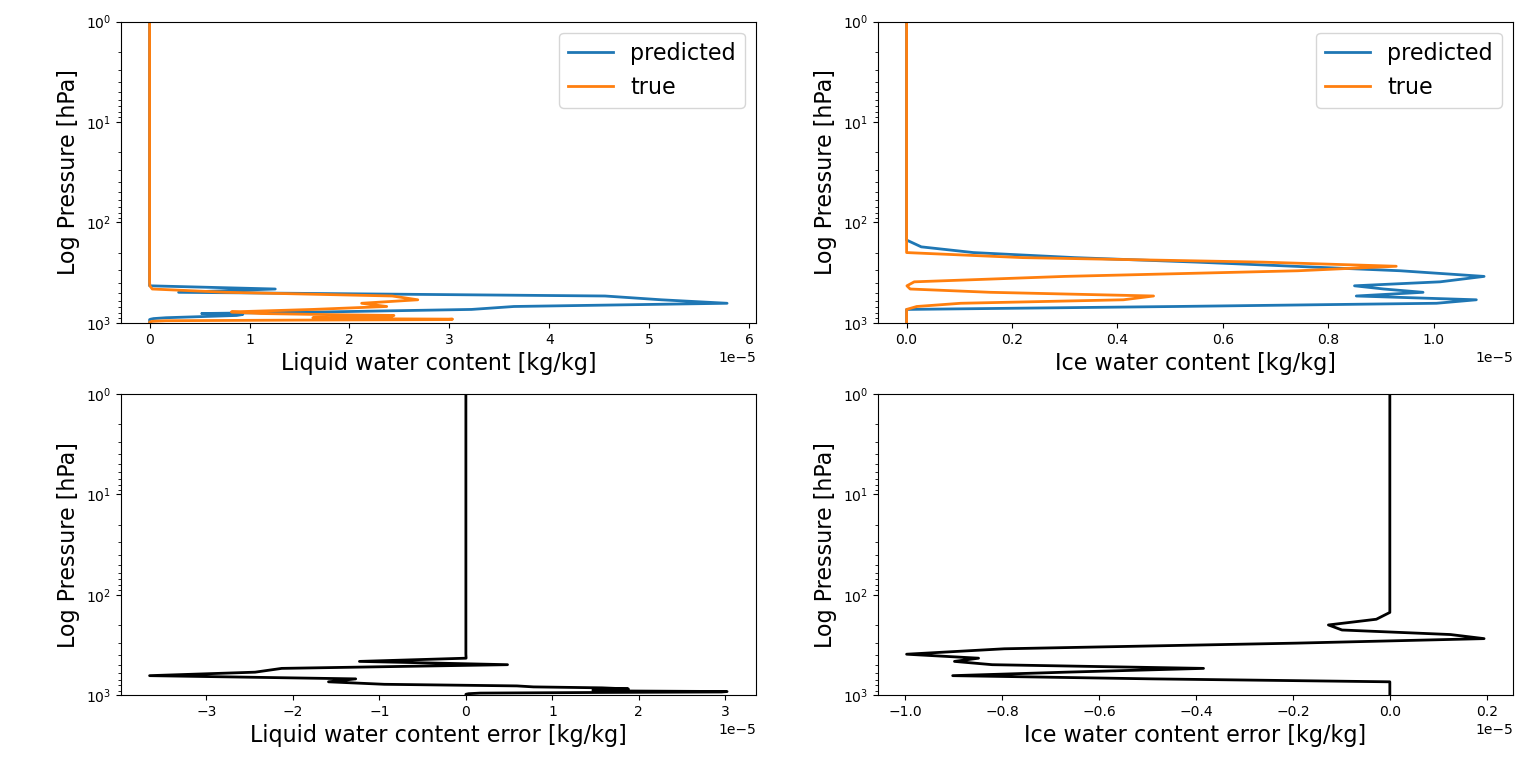}
\caption{Retrieved vertical profiles for cloud contents for the same randomly selected case (case number \num{3102}) after the additional model correction. The top row shows the true profiles (orange) and the retrieved profiles (blue) for cloud specific liquid and ice water contents, while the bottom row displays the corresponding residual errors (black).}\label{fig:case5content2}
\end{center}
\end{figure}

\begin{figure}[H]
\begin{center}
\includegraphics[width=1.
\textwidth]{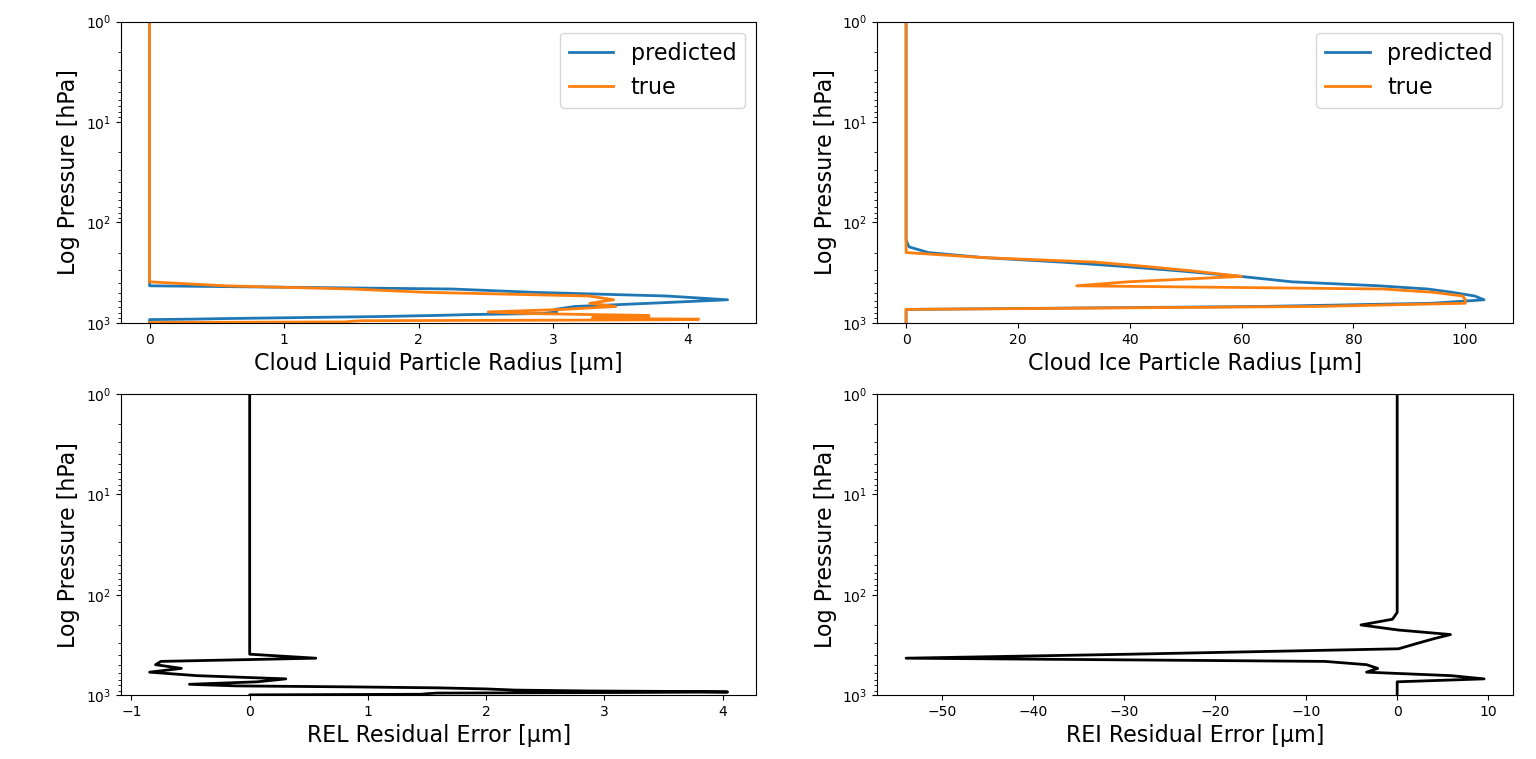}
\caption{Retrieved vertical profiles for cloud particle effective radii for the same randomly selected case (case number\num{3102}) after the additional model correction. The top row shows the true profiles (orange) and the retrieved profiles (blue) for cloud liquid and ice particle effective radii on pressure layers, while the bottom row displays the corresponding residual errors (black).}\label{fig:case5radius2}
\end{center}
\end{figure}

In addition, Figures~\ref{fig:case5atm}, \ref{fig:case5emi}, \ref{fig:case5content2}, and \ref{fig:case5radius2} present the retrieval results for a randomly selected test case (case number~$\num{3102}$). In Figures~\ref{fig:case5atm}, \ref{fig:case5content2}, and \ref{fig:case5radius2}, the first row shows the retrieved profiles (blue) alongside the corresponding reference profiles (orange), while the second row displays the associated residuals (black). Figure~\ref{fig:case5atm} reports, in order, temperature, water vapor, and ozone profiles, with the surface temperature indicated in the caption. Figures~\ref{fig:case5content2} and \ref{fig:case5radius2} show cloud liquid and ice water content, and cloud liquid and ice particle effective radius, respectively. Finally, Figure~\ref{fig:case5emi} presents the retrieved spectral emissivity (dashed blue line) together with the reference profile (orange).

The most important strength of the proposed retrieval framework lies in its computational efficiency. Once trained, the latent twin plus correction is able to invert the spectra for the entire test set in under 2.18 seconds on a standard laptop, corresponding to an average of 0.22 seconds per batch. This remarkable speed highlights the suitability of the method for large-scale applications and near-real-time processing scenarios, where traditional inversion techniques would be computationally prohibitive. The combination of competitive accuracy and extremely low computational cost represents a key advantage of this approach and demonstrates its potential for future operational implementations.

\subsection{Scene Classification Performance}\label{sec:numres_scene}

As an additional validation of the retrieval performance, we analyze the reference total cloud optical depth $\tau$, computed at $900~\si{cm^{-1}}$. This quantity, introduced in~\Cref{sec:tsc} and defined in \Cref{sec:do}, serves as an effective discriminator between clear- and cloudy-sky scenes. In this study, as already mentioned, a threshold value of $\tau = 0.03$ is adopted to classify scenes as clear ($\tau \le 0.03$) or cloudy ($\tau > 0.03$). FORUM simulations~\cite{sgheri2022} indicate that clouds with optical depths below this threshold are generally undetectable.

Using this criterion, we examine the composition of the training and test datasets. In the training set, $\num{3743}$ cases out of $\num{27000}$ (approximately $13.86\%$) are classified as clear-sky. Similarly, in the test set, $685$ cases out of $\num{4862}$ (approximately $14.09\%$) are clear-sky, indicating a consistent proportion between the two datasets. The fraction of clear-sky scenes is lower than the global average of approximately $33\%$~\cite{king2013}. This discrepancy arises because, as discussed previously, cloud fraction information is discarded at the layer level, and each ERA5 tile is instead characterized by its cloudiest atmospheric column. 

\begin{figure}[H]
\centering
\includegraphics[width=\textwidth]{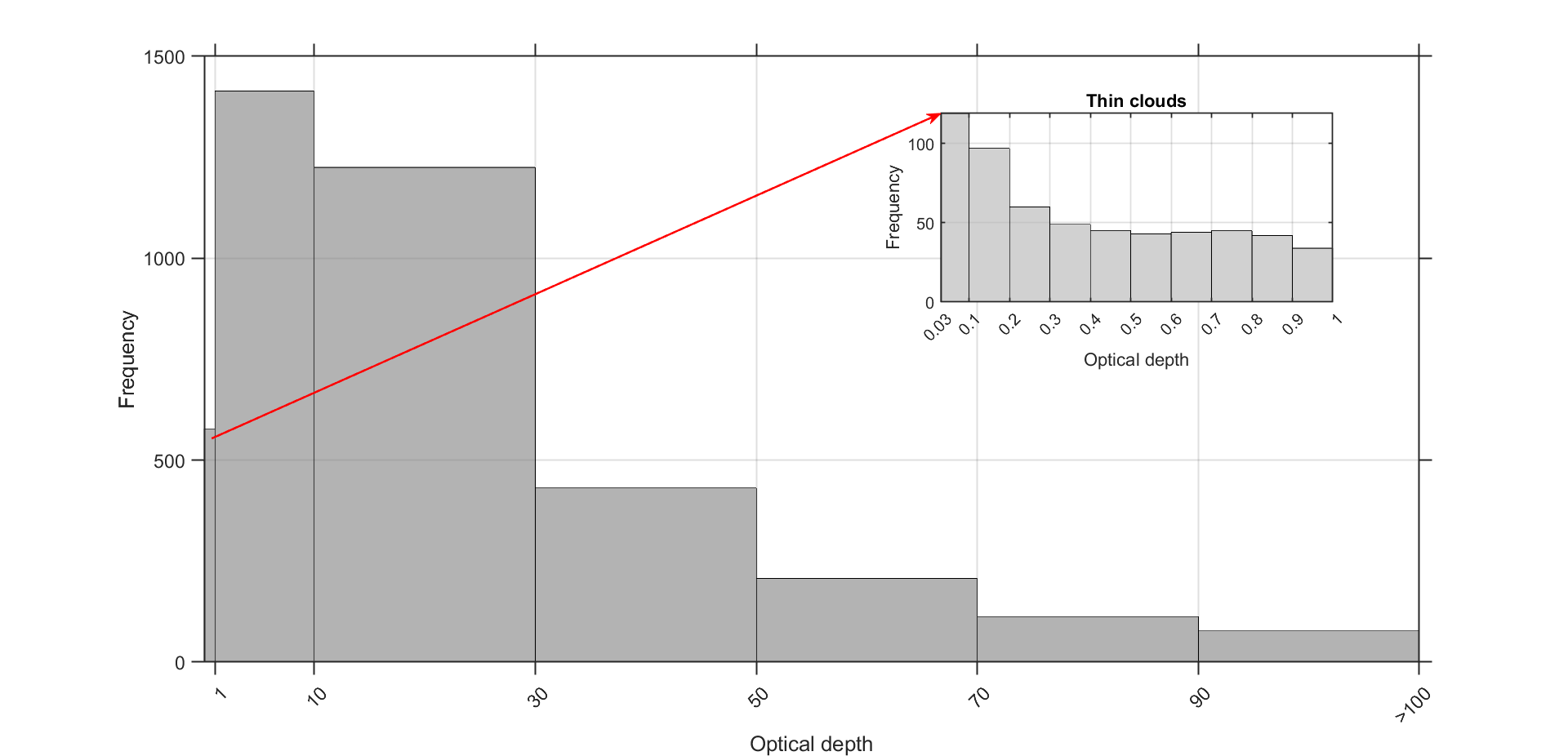}
\caption{Distribution of cloudy cases in the test set as a function of the total cloud optical depth. The inset highlights the low--optical-depth range from $0.03$ (the threshold used to define clear-sky conditions) to $1$, providing a detailed view of the thin-cloud regime in the test set.}
\label{fig:hist}
\end{figure}

Focusing on the test set and excluding clear-sky cases, the distribution of cloudy scenes is shown in the histogram in Figure~\ref{fig:hist}.

We next compare these reference values with the retrieved ones. The retrieved total cloud optical depth, denoted by $\tau^{\text{pred}}$, is not a direct output of the retrieval architecture. Instead, it is analytically derived a posteriori from the retrieved cloud variables using the formulation in \eqref{eq:tau}, by summing the contributions from the liquid and ice components, as described in Section~\ref{sec:do}.

\begin{figure}[H]
\begin{center}
\includegraphics[width=1.
\textwidth]{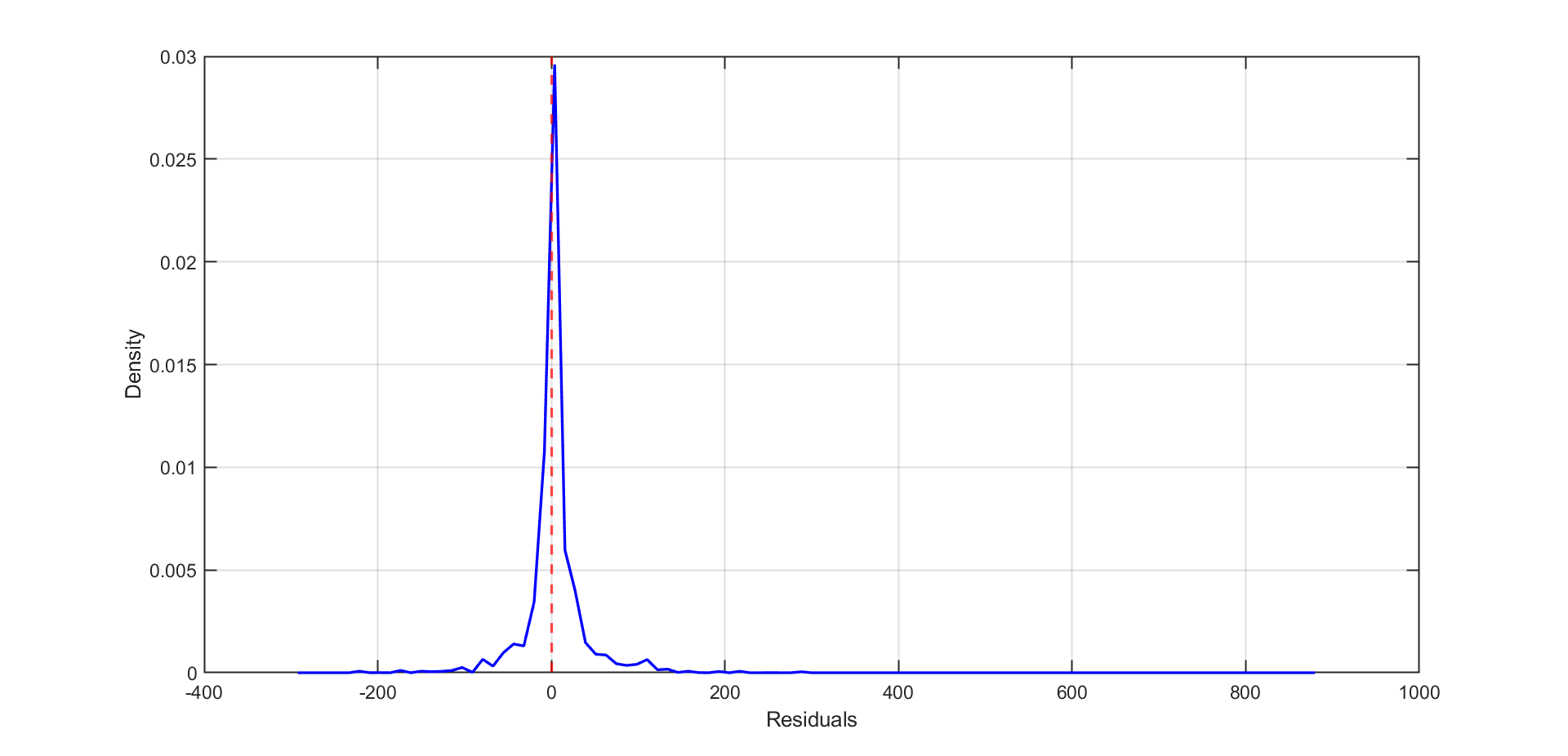}
\caption{Distribution of residuals $\tau^{\text{pred}} - \tau$ in the test set, shown through a kernel density estimate. The vertical dashed red line at zero represents the ideal case of no bias.}\label{fig:residuals}
\end{center}
\end{figure}

Figure~\ref{fig:residuals} shows the distribution of the residuals $\tau^{\text{pred}} - \tau$. The distribution is represented using a kernel density estimate, which provides a smooth approximation of the residual frequency. A vertical dashed line at zero indicates the ideal unbiased case, enabling a visual assessment of potential systematic deviations. This analysis provides complementary insight into the reliability and consistency of the retrievals across a wide range of total cloud optical depths.

\begin{figure}[H]
\begin{center}
\includegraphics[width=1.1
\textwidth]{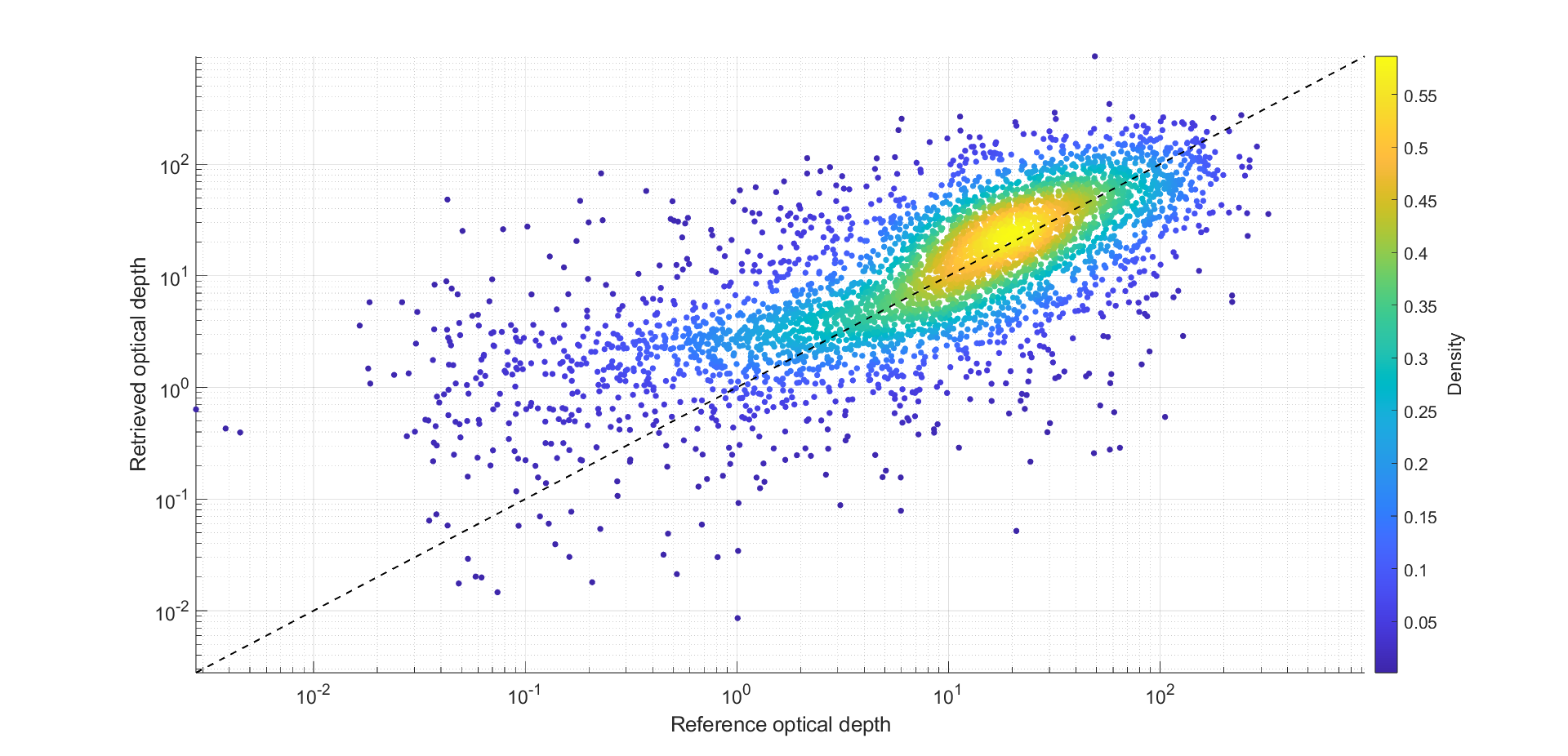}
\caption{Scatterplot of retrieved versus reference total cloud optical depth for the entire test set. Each dot represents a single scene, while the dashed line corresponds to the ideal one-to-one relationship. Values are shown on a logarithmic scale and the color indicates the density of points, as represented by the colorbar on the right.}\label{fig:scatterplot}
\end{center}
\end{figure}

Using the same metrics introduced above, we obtain on the entire test set an $\text{MBE} = 3.10$ and an $\text{MAE} = 17.17$. The scatterplot in Figure~\ref{fig:scatterplot} illustrates the overall agreement between retrieved and reference values across the test set. Total cloud optical depth values are shown on a logarithmic scale to better visualize the full dynamic range. Colors highlight regions where points are more densely clustered.

As shown in Figure~\ref{fig:residuals}, the retrieval of the total cloud optical depth is not exact. However, as demonstrated in \cite{sgheri2022}, optical depths below approximately $0.1$ induce spectral perturbations that are comparable in magnitude to the instrument noise. Conversely, for optically thick clouds ($\tau \gtrsim 10$), further increases in optical depth have only a marginal impact on the observed spectrum, since such clouds effectively block the radiance originating from below the cloud top. As a result, pointwise errors in the retrieved optical depth are not unexpected in these regimes, and the residual distribution should be interpreted with this inherent saturation behavior in mind.

Despite these limitations, the Spearman rank correlation coefficient computed over the test set reaches $0.84$, indicating that the retrieval reliably captures the monotonic relationship between predicted and reference optical depth values, even when absolute deviations are present.

\begin{figure}[H]
\centering
\includegraphics[width=.6\textwidth]{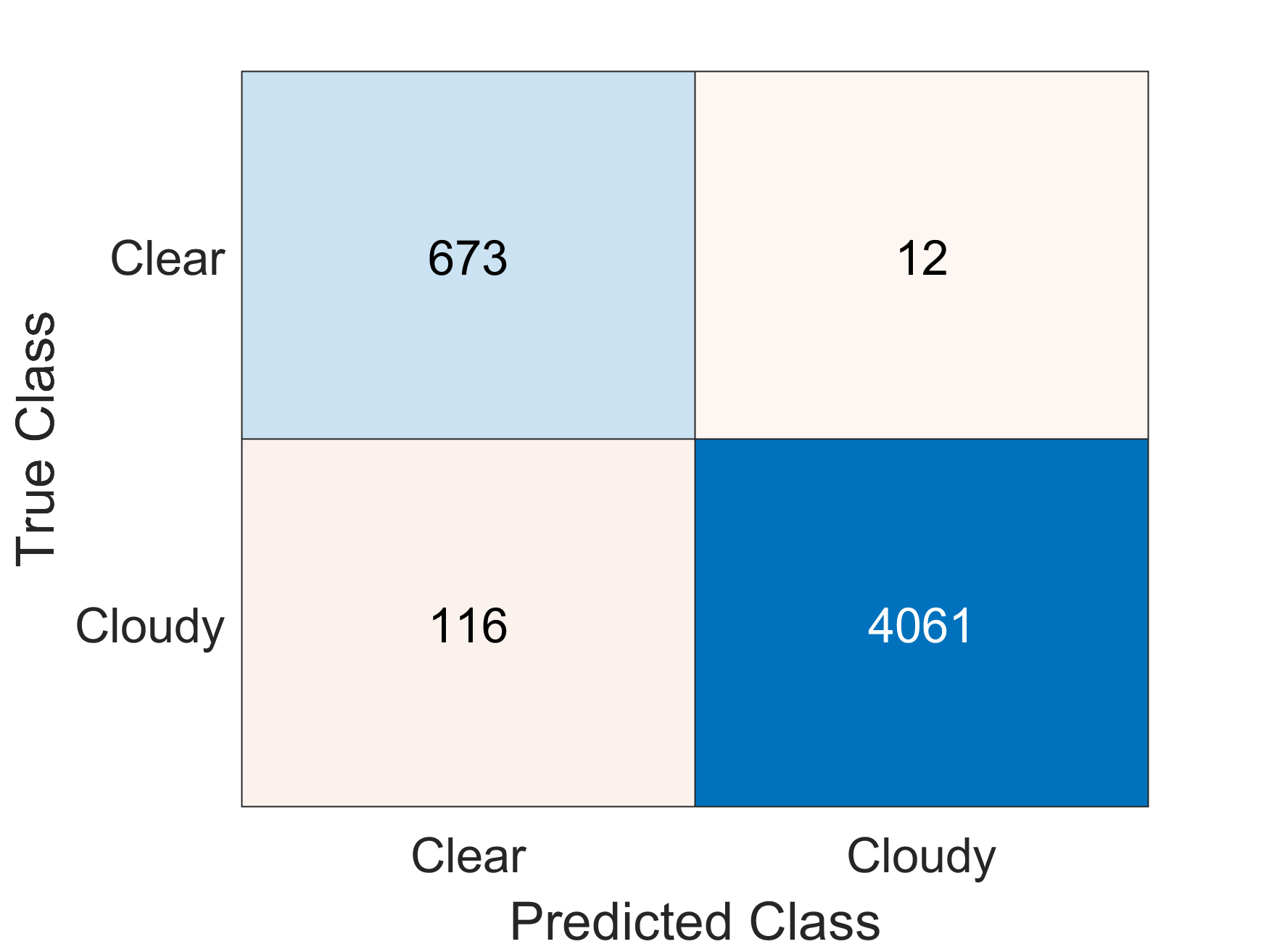}
\caption{Confusion matrix for the classification of clear and cloudy scenes in the test set. Rows correspond to the true classes and columns to the predicted classes. Overall, the classifier correctly identifies $97.37\%$ of the cases. Class-wise accuracies are $98.2\%$ for clear-sky scenes and $97.2\%$ for cloudy scenes.}
\label{fig:confmat}
\end{figure}

The primary strength of the proposed architecture lies in its classification capability rather than in precise regression of optical depth. When used to discriminate between clear and cloudy scenes, the method achieves an overall accuracy of $97.37\%$ on the test set. The corresponding confusion matrix is shown in Figure~\ref{fig:confmat}, where the dominant diagonal entries indicate the high fraction of correctly classified cases. In particular, $98.2\%$ of clear-sky scenes and $97.2\%$ of cloudy scenes are correctly identified.

To the best of our knowledge, there are no directly comparable studies on scene classification based on FORUM-simulated measurements using deep learning approaches. While the cloud detection algorithm embedded in the FORUM End-to-End simulator~\cite{sgheri2022, maestri2019} showed promising results, it was evaluated on a relatively limited test set and does not rely on a deep learning framework. As a comparison, another important infrared interferometer is the Infrared Atmospheric Sounding Interferometer (IASI) \cite{SIMEONI1997}. In this context, Whitburn et al.~\cite{whitburn2022} reported an agreement of $87\%$ between a supervised neural network and the operational IASI Level~2 cloud product using a subset of 45 spectral channels. Higher agreement rates have been achieved for IASI measurements by restricting the analysis to specific geographical regions; for example, Mastro et al.~\cite{mastro2020} reported a $93\%$ agreement over Eastern Europe and tropical regions. In comparison, the results presented here demonstrate that the proposed latent twin framework can achieve competitive, and in some cases superior, classification performance using FORUM-like broadband spectral information over a global dataset.

Furthermore, if we refine the classification into three categories, clear-sky ($\tau \leq 0.03$), thin clouds ($0.03 < \tau \leq 1$), and thick clouds ($\tau > 1$), the percentage of correctly recognized cases in the test set is $87.70\%$.

Finally, to assess whether the reconstructed profiles preserve the vertical structure of the clouds, we perform a Fourier sine decomposition of both the true profiles $\textbf{c}_{\text{liq}}$, $\textbf{c}_{\text{ice}}$, $\textbf{r}_{\text{liq}}$, $\textbf{r}_{\text{ice}}$ and the corresponding predicted ones, denoted with an asterisk.
The goal is not to analyze the content per se, but to verify whether the model captures the same spatial vertical variability patterns as the reference data.
Given a profile $\textbf{c}_{\text{liq}}$, defined at $N$ pressure layers, the coefficients of the sine series, representing the contribution of each harmonic to the profile structure, are computed as
\begin{equation}
    b_k \approx \frac{2}{p^N - p^1} \sum_{i=1}^N \omega^i c^i_{\text{liq}} \sin\Bigg(\frac{k\pi (p^i-p^1)}{p^N-p^1}\Bigg),
\end{equation}
where $k=1,\dots,K$ is the harmonic index, $\omega^i$ are trapezoidal integration weights
$$ \omega^i =
\begin{cases}
    \frac12 (p^2-p^1), & \text{for } i=1, \\
    \frac12 (p^{i+1}-p^{i-1}, & \text{for } i=2, \dots, N-1, \\
    \frac12 (p^N-p^{N-1}), & \text{for } i=N.
\end{cases}
$$
To remove the amplitude dependence and highlight the similarity in the vertical structure of the profiles, the Fourier coefficients are normalized by their maximum absolute value:

$$\textbf{b} = [b_1, \ldots, b_K ],  \qquad \tilde{\textbf{b}} = \frac{\textbf{b}}{\max_i |{b_i}|}.$$

Similarly, we compute the Fourier coefficients $\mathbf{b}^*$ for the predicted variable $\mathbf{c}_{\text{liq}}^*.$

Finally, for each test case, we compute the cosine similarity between the true and reconstructed coefficient vectors $\textbf{b}$ and $\textbf{b}^*$ as:
\begin{equation}
    \cos_{\text{sim}} \beta= \frac{(\tilde{\mathbf{b}}^{*})^\top \cdot \tilde{\mathbf{b}}}{\|\tilde{\mathbf{b}}^*\| \cdot \|\tilde{\mathbf{b}}\|}
\end{equation}
This metric quantifies the degree of alignment between the spectral representations of the true and predicted profiles, with values close to 0 indicating no similarity and values close to 1 denoting an identical structural shape. The same analysis is carried out for the remaining three variables.

Across the test dataset, the cosine similarity averages around 0.75 for water-cloud variables and 0.67 for ice-cloud variables, indicating that the model correctly reproduces the vertical position and general structure of cloud layers, even when the amplitude of the retrieved profiles is not perfectly reconstructed.

\subsection{Radiance Reconstruction} \label{sec:numres_chi2}
In this subsection we apply the radiative transfer to the reconstructed profiles in order to evaluate the error obtained on the radiance. We use the reduced spectral $\chi^2_r$ statistics, defined as:
$$
    \chi^2_r = \frac{\big(\bfy-\bfF^\to(\bfx)\big)\t\bfS_\bfy^{-1}\big(\bfy-\bfF^\to(\bfx)\big)}{q}
$$
where $\bfS_\bfy$ is the covariance matrix of the measurements that can be calculated from the known apodization kernel and NESR level. The expected value of $\chi^2_r$ for a spectrum that is compatible with the instrumental noise is 1, and for a spectrum reconstructed after minimization is $(q - n)/q \simeq 0.9$. For all points in the dataset, applying the forward model to the true atmospheric state produces spectra with $\chi^2_r \equiv 1$ when evaluated against the noisy spectra.

In order to improve the agreement between the model and reconstructed spectra for cases classified as clear, we applied a simple post-processing step. Specifically, we set the reconstructed surface temperature $T^0$ as the solution of the equation $s^{900} = \epsilon^{900} B(T^0)$, where $B$ is the Planck function, $s^{900}$ is the radiance at $900$~\si{cm^{-1}}, and $\epsilon^{900}$ is the reconstructed emissivity at the same wavenumber. This approach assumes negligible atmospheric absorption at $900$~\si{cm^{-1}}, which lies within the atmospheric window. This simple procedure significantly improves the agreement of the reconstructed spectra.

\begin{figure}[H]
\begin{center}
\includegraphics[width=0.5
\textwidth]{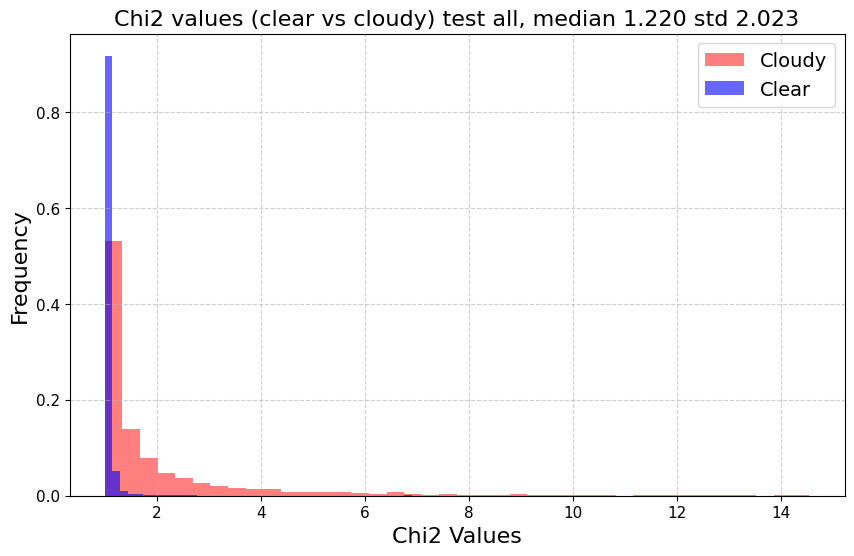}
\caption{$\chi^2_r$ obtained by applying the stretching and de-stretching procedure to the true atmospheric states over the entire test set. Clear-sky cases are shown in blue, while cloudy cases are shown in red. The median value and the corresponding standard deviation are reported in the figure title.}\label{fig:chi2_hist_true}
\end{center}
\end{figure}

First of all, in our results we observe that the $\chi^2_r$ error is strongly influenced by the stretching procedure, described in \Cref{sec:do} and previously adopted in \cite{sgattoni2025}, which is applied as a preprocessing step to obtain vertically uniform data with consistent dimensionality. To illustrate this effect, Figure~\ref{fig:chi2_hist_true} shows the $\chi^2_r$ obtained by applying only the stretching and de-stretching procedure to the true dataset, demonstrating that this operation alone already introduces a deviation from the expected value discussed above.

\begin{figure}[htbp]
\centering
\begin{minipage}{0.49\textwidth}
  \centering
  \includegraphics[width=\linewidth]{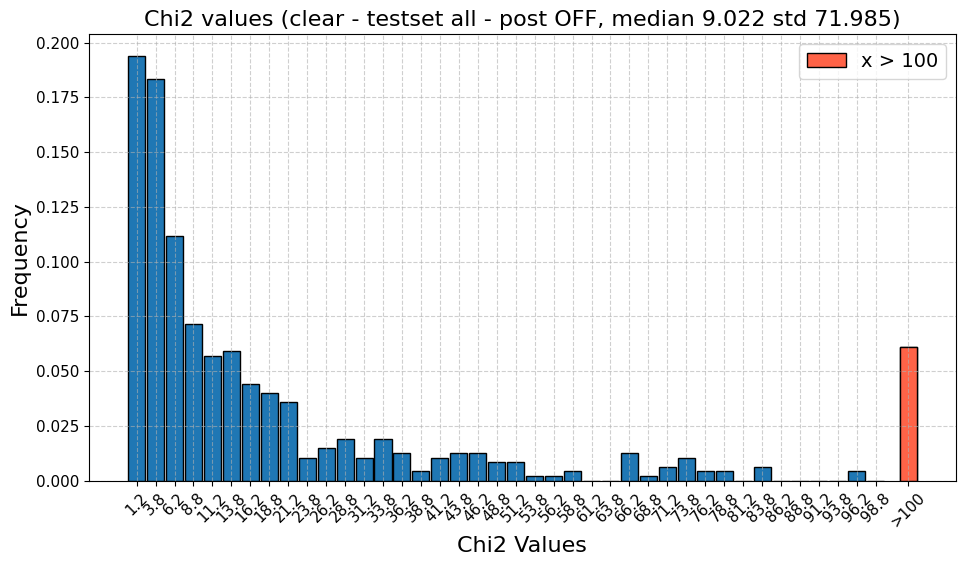}
\end{minipage}
\hfill
\begin{minipage}{0.49\textwidth}
  \centering
  \includegraphics[width=\linewidth]{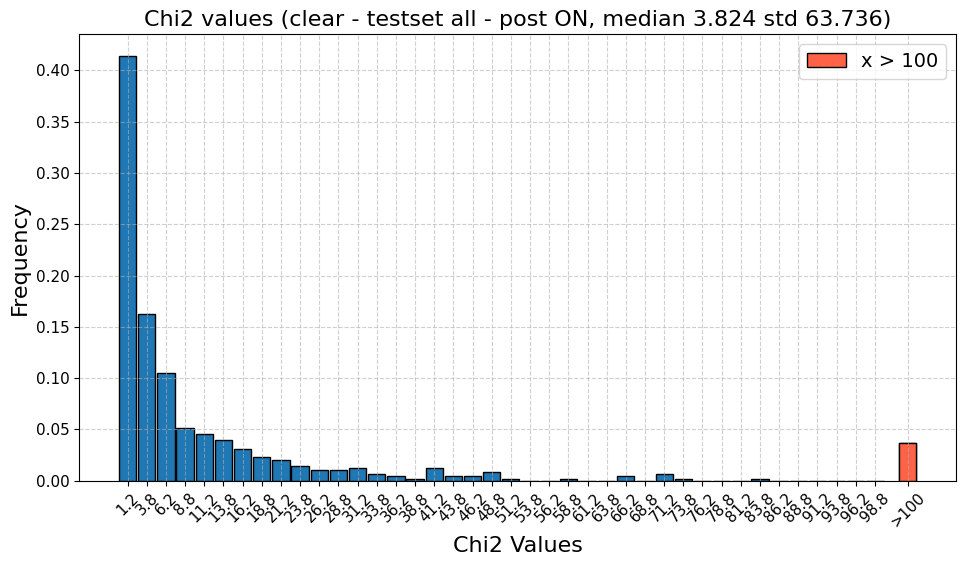}
\end{minipage}

\vspace{0.2cm}

\begin{minipage}{0.49\textwidth}
  \centering
  \includegraphics[width=\linewidth]{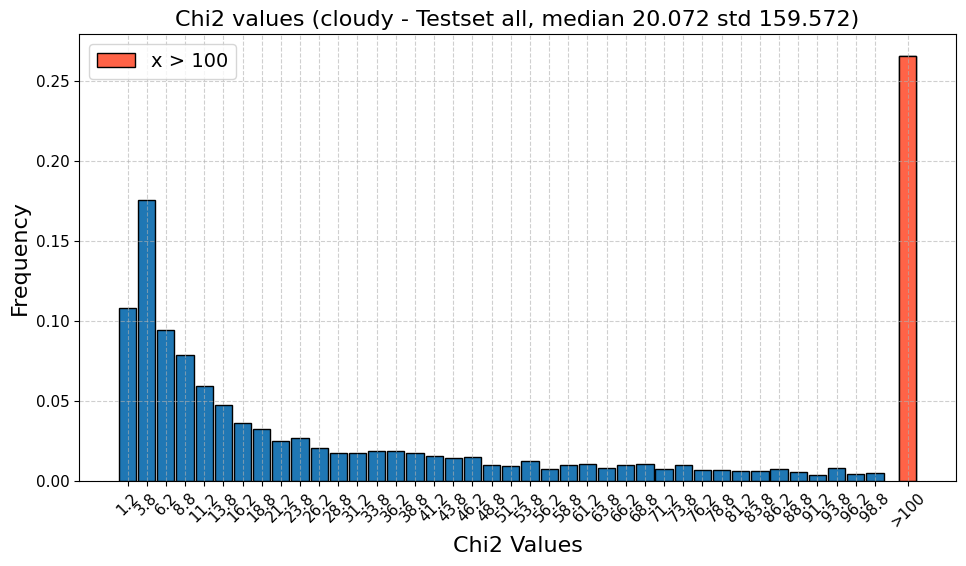}
\end{minipage}
\hfill
\begin{minipage}{0.49\textwidth}
  \centering
  \includegraphics[width=\linewidth]{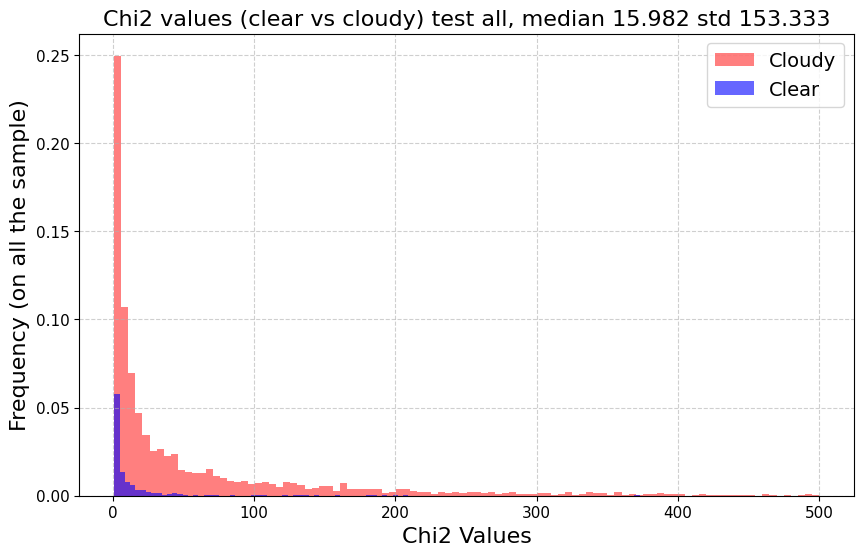}
\end{minipage}

\caption{$\chi^2_r$ obtained after reconstruction. The top row displays clear-sky cases, without (left) and with (right) the postprocessor, while the bottom-left panel shows cloudy cases only, and the bottom-right panel shows the combined histogram for all test cases. All quantitative information, including the median value and its standard deviation, is reported in the figure title. In the bottom-right panel, clear-sky cases are shown in blue and cloudy cases in red. In the other panels, red only highlights higher $\chi^2_r$ values.}\label{fig:chi2_all}
\end{figure}

In Figure~\ref{fig:chi2_all} we show the results of the reconstruction, which therefore also include the error introduced by the stretching procedure discussed above.  In the top row, we present the results for the cases recognized as clear by the system: the left panel shows the reconstruction without the postprocessor, and the right panel shows the reconstruction with the postprocessor. In the bottom row, the left panel displays the results for the cases recognized as cloudy, while the right panel shows the combined histogram for all cases in the test set.

\begin{figure}[H]
\begin{center}
\includegraphics[width=0.5
\textwidth]{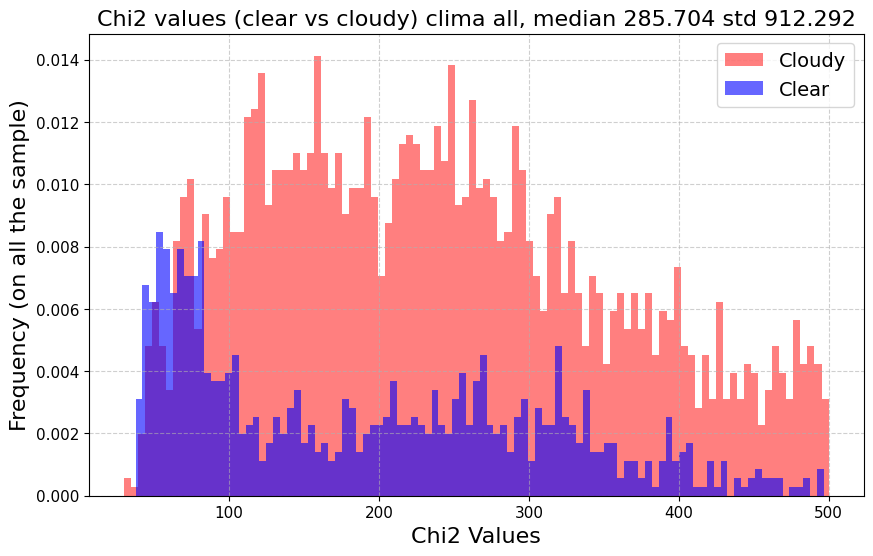}
\caption{$\chi^2_r$ obtained using climatological a priori for the full test set. All quantitative information, including the median and its standard deviation, is reported in the figure title. Clear-sky cases are shown in blue and cloudy cases in red}\label{fig:chi2_clima}
\end{center}
\end{figure}

It is evident that for most cases the $\chi^2_r$ is not satisfactory for a real retrieval, still the values obtained are not a bad estimate to be used as an initial guess or an a priori for a refinement retrieval. Just for comparison, in Figure~\ref{fig:chi2_clima} we show the combined $\chi^2_r$ distribution obtained perturbing the atmospheric state according to atmospheric variability, obtained from the IG2 estimates for clear sky variables, and determined from the dataset itself for the cloud variables. Even if we suppose that clear and cloudy cases are correctly identified, the $\chi^2_r$ distribution is much worse than our reconstruction. Overall, these results show that our reconstruction provides a significantly better starting point than simply using the climatological a priori.

\section{Conclusion} \label{sec:concl}

In this work, we introduced a data-driven, physics-correcting retrieval framework for FORUM all-sky observations based on the \emph{latent twins} framework. By coupling paired autoencoders for atmospheric states and radiance spectra through trainable latent-space mappings, the proposed approach learns a surrogate inverse operator that is both computationally efficient and constrained to a physically plausible manifold. This design has the potential to enable fast, stable retrievals for a highly ill-posed and high-dimensional inverse problem, while retaining a clear interpretation in terms of forward and inverse radiative transfer surrogates.

In this study, the method is applied to FORUM specifications and ERA5 reanalysis data, but it is readily extendable to other infrared sounders.  

The proposed framework shows promise along two main directions. First, it demonstrates strong performance in distinguishing clear-sky from cloudy conditions, achieving an overall classification accuracy of $97.37\%$ over a global dataset. The results are consistent with, and complementary to, previous studies based on FORUM-simulated measurements, such as Maestri et al.~\cite{maestri2019}, one of the few available references using FORUM simulations.
Second, the framework provides a viable basis for the retrieval of atmospheric properties under all-sky conditions. In particular, it enables the retrieval of temperature, water vapor, ozone, and surface emissivity profiles with minimal bias relative to their natural variability.  
Retrieval of cloud-related variables remains inherently more challenging; however, the method captures key cloud features, including the vertical location of cloud layers and their phase, and provides informative estimates of cloud ice and liquid water content as well as cloud particle effective radius. These results indicate that the framework offers a promising basis for future improvements in cloud property retrieval. Moreover, the model-consistency correction introduced for cloud variables resolves physically inconsistent states, such as nonzero cloud particle effective radii in cloud-free layers, at negligible computational cost.



A key strength of the proposed approach is its computational efficiency. Once trained, the latent twin model performs retrievals orders of magnitude faster than full-physics variational inversion methods, enabling near-real-time inference on standard hardware. Radiance reconstruction experiments further show that the retrieved atmospheric states provide substantially improved spectral consistency compared with climatological priors, making them well suited as initial guesses, adaptive priors, or proposal states for downstream physics-based refinement and data assimilation workflows.

Several important extensions remain for future work. These include training on more diverse seasonal and diurnal conditions, incorporating additional trace gases and uncertainty-aware formulations, and integrating the latent twin retrieval into hybrid inversion schemes that combine learned surrogates with iterative physics-based solvers. Extending the framework to explicitly quantify uncertainty, for example through Bayesian latent twins or ensemble-based formulations, is another promising direction. Finally, application to real observations from current infrared sounders and, ultimately, to FORUM flight data will be essential to fully assess robustness and generalization.

Overall, this work shows that latent twins provide a powerful and flexible framework for fast, physically informed inversion of complex remote-sensing measurements. By bridging modern machine learning with established radiative transfer modeling, the proposed approach opens new pathways for operational retrievals, large-scale data exploitation, and future assimilation of far-infrared observations into weather and climate models.

\section{Acknowledgments}
INdAM-GNCS supported the first author under Bando di concorso a n.45 mensilità di Borse di studio per l’estero A.A. 2023-2024. NRRP (National Recovery and Resilience Plan) supported the first and second authors under the project EMM (Earth-Moon-Mars, Mission 4, Component 2, Investment 3.1, Project IR000038, CUP: C53C22000870006). ASI (Italian Space Agency) supported the second author under the project CASIA (CAirt and Sinergy with IAsi-ng), CUP: F93C23000430001.

\section{Data availability}
The training and test datasets used in this study are available for public access at \url{https://doi.org/10.5281/zenodo.17969539} \cite{sgattonizenodo2025}.

\printbibliography

\begin{appendix}

\section{Forward Model}  \label{sec:app1}
The forward mapping $\mathbf{F}^\to(\mathbf{x})$ represents a combination of the radiative transfer equation and the instrumental effects. Assuming a horizontal and azimuthal symmetry for the problem, the radiative transfer equation can be written as the following initial value problem:
\begin{equation}\label{eq:rt1}
\begin{cases}
    \mu \frac{d I(\tau,\mu)}{d \tau} = I(\tau,\mu) -\frac{\omega(\tau)}{2}\int^{1}_{-1}P(\tau,\mu,\mu')I(\tau,\mu')d\mu' - \left[ 1-\omega(\tau) \right] B(\tau),\\ 
    I(\tau_0,\mu)=I_{0}(\mu),
    \end{cases}
\end{equation}

where $\mu$ denotes the cosine of the zenith angle, $\tau$ the integrated total cloud optical depth from the top of the atmosphere to the level of interest, $\omega$ the single scattering albedo of the layer at level $\tau$, $B$ the Planck function at that level, and $P(\tau,\mu,\mu')$ the azimuthally averaged scattering phase function that describes scattering events for radiation entering the layer at angle $\mu'$ and exiting at $\mu$.
The integral term in equation \ref{eq:rt1}, describing the multiple-scattering interactions, substantially increases the computational cost of the solution. This is mainly due to the increased dimensionality of the problem in solvers such as the DIScrete Ordinate Radiative Transfer (DISORT) method \cite{stamnes1988,stamnes2000,sgheri2018}. A possible solution for mitigating the computational burden is to avoid the direct calculation of the multiple scattering term by scaling the absorption optical depth of the cloud.
The fast radiative transfer code $\sigma$-IASI/F2N \cite{masiello2024b} exploits scaling methodologies to accelerate radiative transfer calculations while maintaining solution accuracy. Specifically, the code implements the Chou approximation \cite{chou1999} and the Tang adjustment routine \cite{tang2018}.
In this work, to generate the training and test sets, the Chou solution is considered. This scaling methodology allows considering an apparent absorption optical depth $\tilde{\tau}$, defined as

\begin{equation}
    \tilde{\tau} = \tau \bigl( (1-\omega) + \omega b \bigr)
\end{equation}
where $b$ is called back-scattering coefficient and describes the mean fraction of radiation back-scattered from the medium. After this scaling, the radiative transfer equation can be written as a Schwarzschild-like form, and the radiative transfer problem becomes

\begin{equation}\label{eq:rt2}
\begin{cases}
    \mu \frac{d I(\tilde{\tau},\mu)}{d \tilde{\tau}} = I(\tilde{\tau},\mu) - \left[ 1-\omega(\tilde{\tau}) \right] B(\tilde{\tau}),\\ 
    I(\tilde{\tau}_0,\mu)=I_{0}(\mu),
    \end{cases}
\end{equation}

Finally, the instrumental effects are modeled by convolving the pseudo-monochromatic radiances produced by $\sigma$-IASI/F2N with the theoretical FORUM instrumental response function.



\section{Input Preprocessing for Radiance Reconstruction} \label{sec:app2}

\begin{figure}[H]
  \centering
  \includegraphics[width=0.5\textwidth]{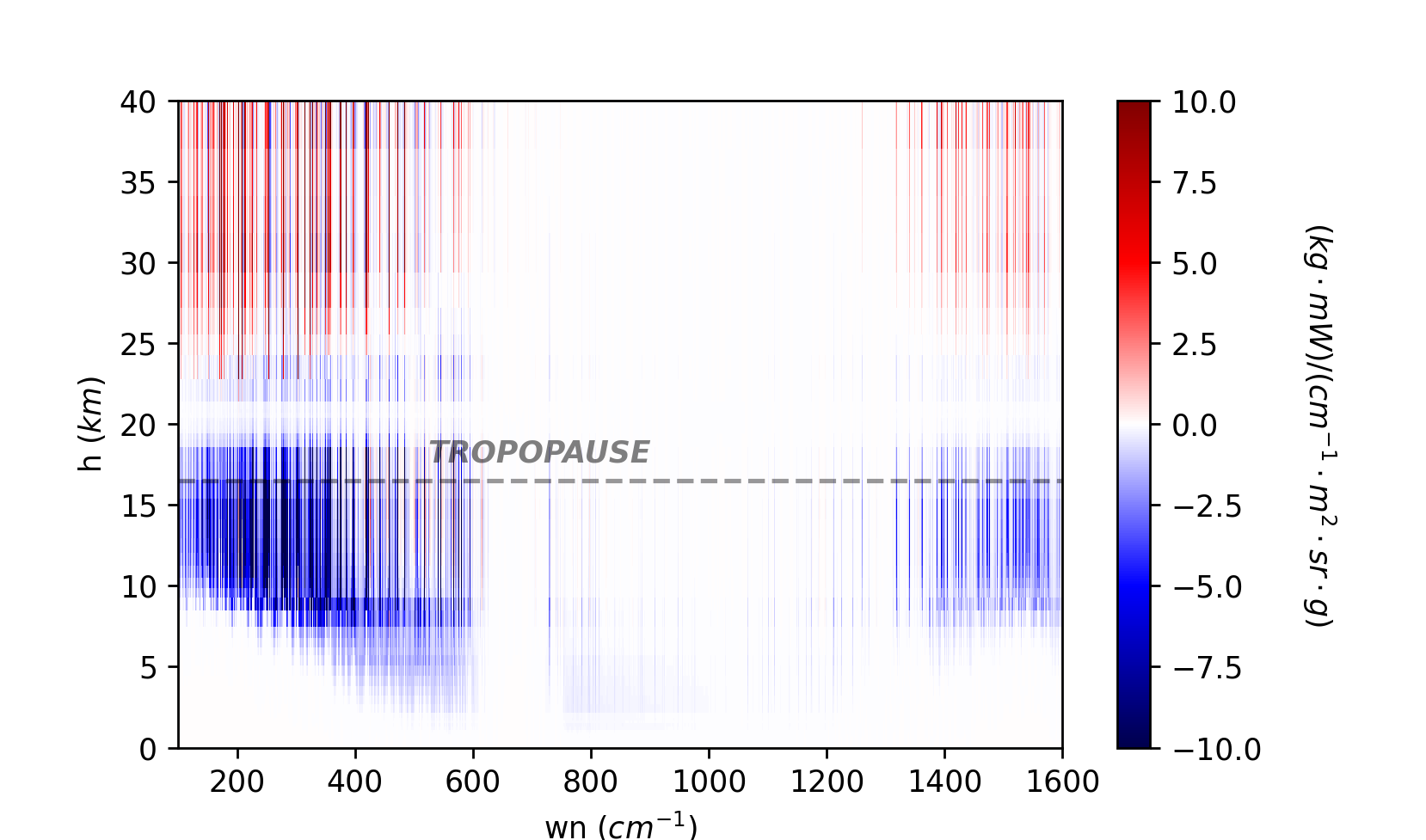}%
  \hfill
  \includegraphics[width=0.5\textwidth]{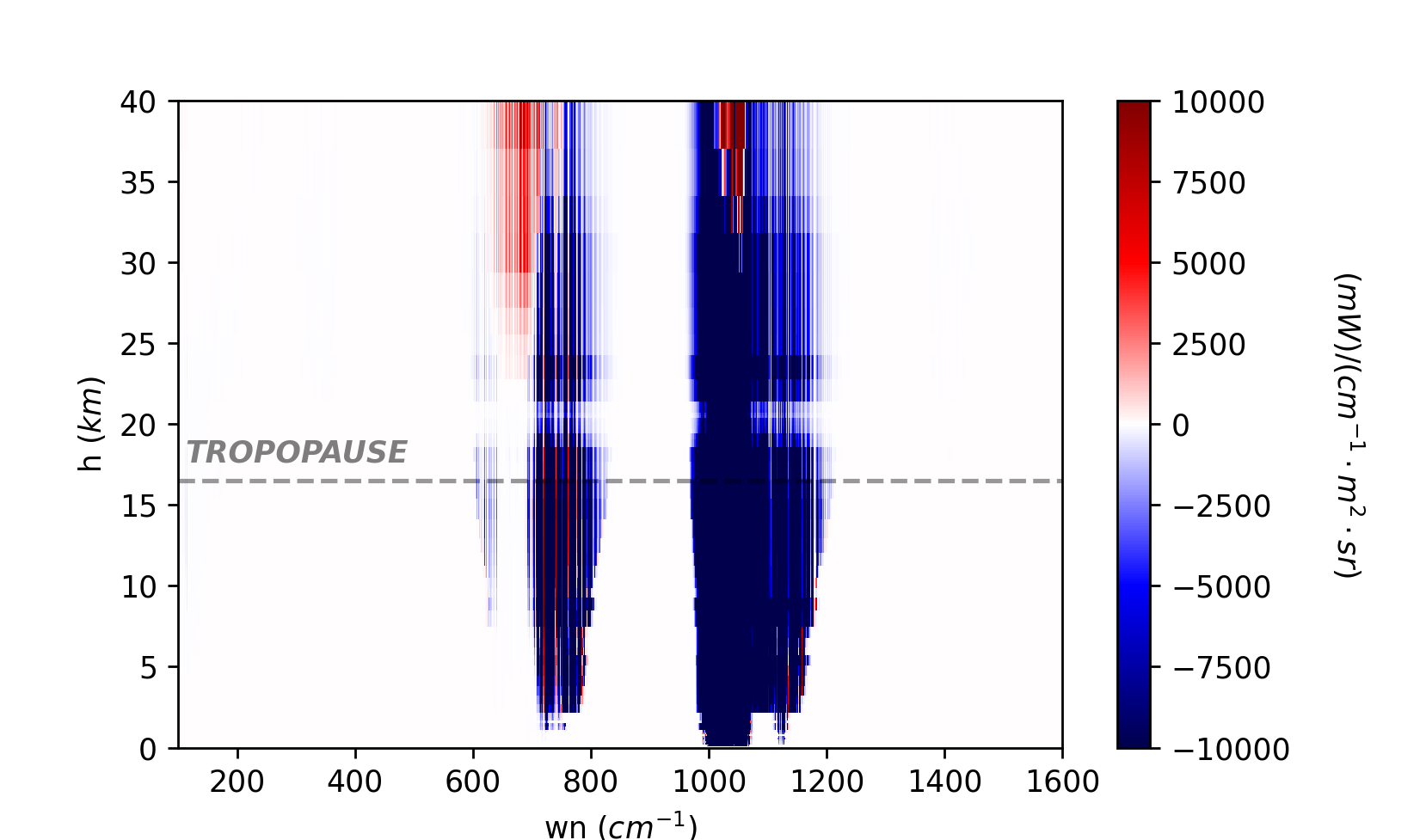}
	\caption{$\sigma$-IASI/F2N Jacobians with respect to the water vapor concentration (left panel) and ozone concentration (right panel). These quantities are calculated considering the US standard atmosphere and successively convolved to FORUM resolution.}\label{covariance_matrix_high_cld}
\end{figure}

Studying how radiance responds to perturbations in atmospheric profiles provides valuable insight into the most appropriate pre‑processing to apply to the input variables of the inversion model. Figure \ref{covariance_matrix_high_cld} shows the derivatives of the forward model with respect to the water vapor concentration $\frac{\partial s}{\partial w_{\text{vap}}}$, and ozone concentration $\frac{\partial s}{\partial o}$, respectively. These quantities are directly computed by $\sigma$-IASI/F2N considering the US standard atmosphere as input scenario.

The radiance shows a strong sensitivity to water vapor variations across the whole vertical profile, with a similar behavior observed for the ozone channels. Accurately reconstructing the vertical distributions of water vapor and ozone throughout the atmosphere is therefore essential to ensure consistency between the measured spectrum and the radiance simulated from the retrieved quantities.\\
In this work, the vertical concentration of water vapor and ozone are transformed using the inverse softplus function. Specifically:

\begin{equation}
    \tilde{x} = ln(e^{{x}\cdot c}-1)
\end{equation}

where $x$ is the quantity of interest and $c$ is a scaling constant, which is set to be $c=1$ for water vapor and $c=10^6$ for ozone. This transformation behaves approximately linearly for $x>>1$ and approximately logarithmically for $x<<1$. This reduces the range of magnitudes spanned by the physical quantities and helps the model, during training, to give sufficient importance to regions where the concentration profile is very low, such as stratospheric water vapor and tropospheric ozone.

\end{appendix}

\end{document}